%% file: SIQUPAPER00.tex
\documentclass[nofootinbib,aps,prb,twocolumn,secnumarabic,balancelastpage,amsmath,amssymb,floatfix,superscriptaddress,10pt]{revtex4-2}

\usepackage{xcolor}

\usepackage{bm}
\usepackage{mathtools}

\newcommand{\tr}{\operatorname{Tr}}
\newcommand{\ketbra}[2]{\ket{#1} \hskip -0.8ex \bra{#2}}

\newcommand{\DP}{D'yakonov-Perel' }

\newcommand{\si}[1]{\ce{^{#1}Si}} 
\newcommand{\up}{\uparrow}
\newcommand{\dn}{\downarrow}
\newcommand{\expct}[1]{\left\langle #1 \right\rangle}
\newcommand{\cbox}{{\mathrm{box}}}

\usepackage{amssymb}
\usepackage{amsmath}
\usepackage{amsfonts}
\usepackage{braket}
\usepackage{graphicx}
\usepackage{epstopdf}
\usepackage{epsfig}
\usepackage[toc,page,title,titletoc,header]{appendix}
\usepackage{dsfont,amsthm,amsbsy}
\usepackage{hyperref}
\usepackage[xspace]{ellipsis}
\usepackage{cancel}
\usepackage[version=4]{mhchem}

\DeclareMathOperator{\erfc}{erfc}

\DeclareMathOperator{\var}{var}

\newcommand{\pevcm}{\mathrm{peV\cdot cm}}
\newcommand{\contact}{\text{contact}}
\newcommand{\nucl}{\text{nucl}}

\newcommand{\init}{\text{init}}

\makeatletter
\newcommand{\appendixsubsectionformat}{%
  \begingroup
  \renewcommand{\p@subsection}{\thesection.}
  \renewcommand{\p@subsubsection}{\thesection.\thesubsection.}
}
\makeatother

\newcommand{\depends}[1]{{ #1} } %

\begin{document}

\title{Electrical Interconnects for Silicon Spin Qubits}

\author{Christopher David White}
\email{christopher.d.white117.ctr@us.navy.mil}
\affiliation{Joint Center for Quantum Information and Computer Science, NIST/University of Maryland, College Park, MD 20742, USA}
\affiliation{Center for Computational Materials Science, U.S. Naval Research Laboratory, 
\\
Washington, D.C. 20375, USA}

\author{Anthony Sigillito}
\affiliation{Department of Electrical and System Engineering, University of Pennsylvania, Philadelphia,
Pennsylvania 19104, USA}

\author{Michael J. Gullans}
\email{mgullans@umd.edu}
\affiliation{Joint Center for Quantum Information and Computer Science, NIST/University of Maryland, College Park, MD 20742, USA}

\begin{abstract}
Scalable spin qubit devices will likely require long-range qubit interconnects. We propose to create such an interconnect with a resistive topgate.
The topgate is positively biased, to form a channel between the two dots;
an end-to-end voltage difference across the nanowire results in an electric field
that propels the electron from source dot to target dot.
The electron is momentum-incoherent, but not necessarily spin-incoherent;
we evaluate threats to spin coherence due to spin-orbit coupling, valley physics, and nuclear spin impurities.
We find that spin-orbit coupling is the dominant threat,
but momentum-space motional narrowing due to frequent scattering partially protects the electron,
resulting in characteristic decoherence lengths $\sim$ 15 mm for plausible parameters.
\end{abstract}

\maketitle

\section{Introduction}

Silicon spin qubits \cite{lossQuantumComputationQuantum1998,zwanenburgSiliconQuantumElectronics2013,burkardSemiconductorSpinQubits2023} offer a promising platform for scalable quantum computation, due to existing fabrication infrastructure and the small size of the physical qubits.
Recent experiments have demonstrated high-fidelity state preparation, readout, and single- and two-qubit gates \cite{yonedaQuantumdotSpinQubit2018,xueQuantumLogicSpin2022b,noiriFastUniversalQuantum2022,millsHighFidelityState2022,millsTwoqubitSiliconQuantum2022a}.
But silicon spin qubits require dense control electronics, which constrain both device layouts and high-level properties like qubit connectivity.
Moreover, both NISQ calculations \cite{linkeExperimentalComparisonTwo2017} and many quantum error correcting codes \cite{breuckmannQuantumLowDensityParityCheck2021} benefit from long-range gates.

A large-scale spin qubit device will therefore require not only good qubits,
but also medium and long-range interconnects between qubits
\cite{taylorFaulttolerantArchitectureQuantum2005a,vandersypenInterfacingSpinQubits2017}.
Existing proposals for interconnects include
adiabatic passage through an array of quantum dots
\cite{greentreeCoherentElectronicTransfer2004,greentreeQuantuminformationTransportMultiple2006,buslBipolarSpinBlockade2013,sanchezLongRangeSpinTransfer2014,banSpinEntangledState2019,groenlandStimulatedRamanAdiabatic2020,stefanatosSpeedingAdiabaticPassage2020,gullansCoherentTransportSpin2020a},
bucket-brigade shuttling
\cite{baartSinglespinCCD2016a,flentjeCoherentLongdistanceDisplacement2017,fujitaCoherentShuttleElectronspin2017a,millsShuttlingSingleCharge2019b,mortemousqueEnhancedSpinCoherence2021,yonedaCoherentSpinQubit2021b,noiriShuttlingbasedTwoqubitLogic2022a,zwerverShuttlingElectronSpin2023,vanriggelen-doelmanCoherentSpinQubit2023,smetHighfidelitySinglespinShuttling2024},
conveyor-mode shuttling
\cite{taylorFaulttolerantArchitectureQuantum2005a,langrockBlueprintScalableSpin2023,seidlerConveyormodeSingleelectronShuttling2022a,struckSpinEPRpairSeparationConveyormode2023,xueSiSiGeQuBus2023},
coupling to microwave cavities
\cite{childressMesoscopicCavityQuantum2004,majerCouplingSuperconductingQubits2007,viennotCoherentCouplingSingle2015,miCoherentSpinPhoton2018,samkharadzeStrongSpinphotonCoupling2018,landigCoherentSpinPhoton2018,borjansResonantMicrowavemediatedInteractions2020,harvey-collardCoherentSpinSpinCoupling2022,dijkemaTwoqubitLogicDistant2023},
and more \cite{shiltonHighfrequencySingleelectronTransport1996,jadotDistantSpinEntanglement2021,vandiepenElectronCascadeSpin2021,hetenyiLongdistanceCouplingSpin2022}.
In bucket-brigade interconnects, the electron undergoes coherent adiabatic shuttling through a sequence of dots,
controlled by precise modulations of barrier and accumulator gate potentials.
Conveyor-mode devices avoid some limitations of bucket-brigade shuttling.
Rather than passing through Landau-Zener transitions from dot to dot,
an electron in a conveyor-mode interconnect travels in a moving potential
created by many gates with relatively few control lines;
the electron speed is a few meters per second.
In microwave interconnects, qubits are coupled to a microwave resonator,
leading to effective two-qubit gates between qubits separated by distances up to millimeters.

\begin{figure}[t]
  \includegraphics[width=\columnwidth]{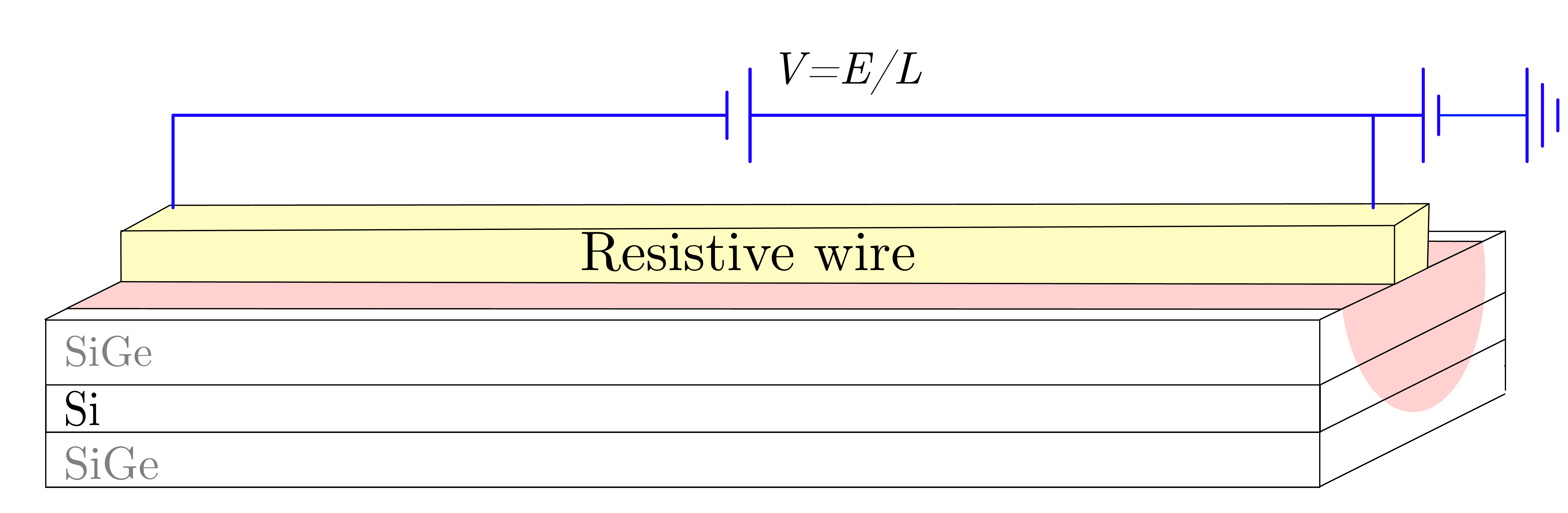} \vskip 0.3cm
  
  \includegraphics[width=0.61\columnwidth]{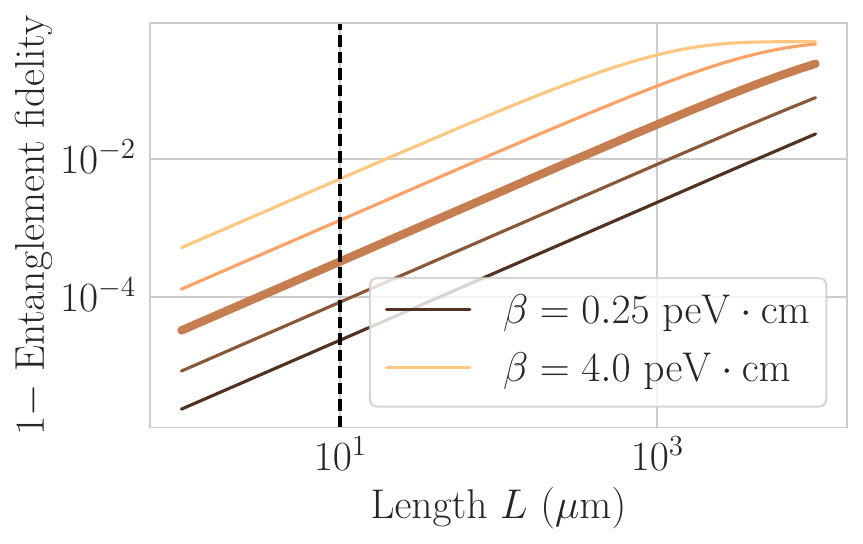}
  \includegraphics[width=0.37\columnwidth]{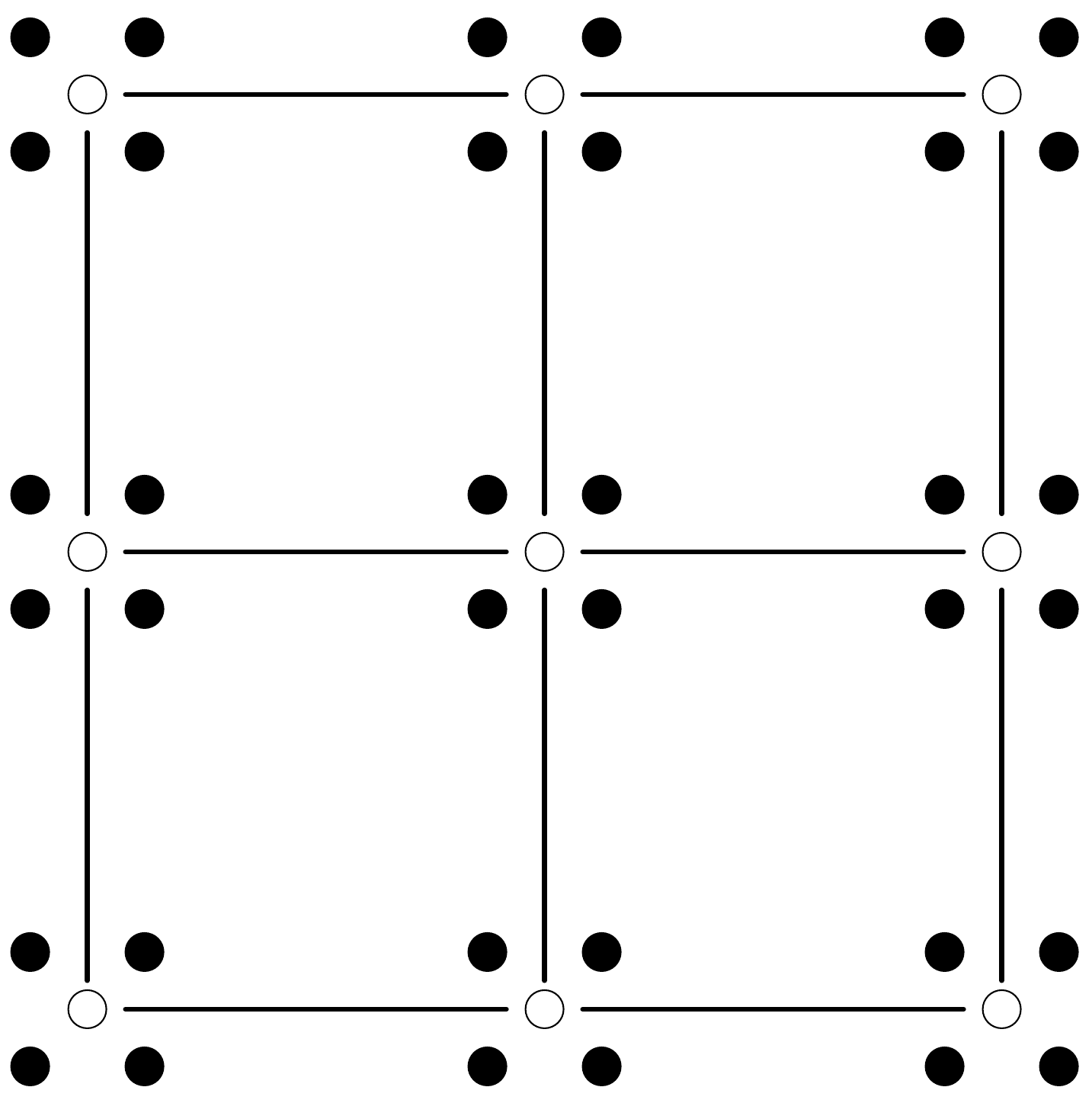}
  
  \caption{\textbf{Top: proposed device}.
  A Si/SiGe heterostructure confines an electron to the thin Si layer.
  A resistive topgate creates a one-dimensional channel in the Si layer and generates an electric field along the channel;
  an electron in the channel undergoes momentum-incoherent but spin-coherent transport down the channel.
  \textbf{Bottom left: entanglement fidelity} of transport across the channel as a function of channel length and spin-orbit interaction.
  Spin-orbit interaction couples the electron's spin degree of freedom to its momentum degree of freedom, which is incoherent due to scattering.
  Momentum-space motional narrowing, caused by frequent scattering, partially protects the spin degree of freedom.
  The bold line shows Dresselhaus spin-orbit coupling $\beta = 1\;\mathrm{peV\cdot cm}$, the value we expect for our devices (cf App.~\ref{app:soc:value}.)
  \textbf{Bottom right:} a potential architecture using interconnects to connect plaquettes of five dots, four filled (filled circles) and one empty (empty circle).
  }
  \label{fig:summary}
\end{figure}

We propose to shuttle electrons between dots separated by distances $L \sim \depends{100}\;\mu$m
using a single nanowire as a resistive topgate (Fig.~\ref{fig:summary}).
The topgate is positively biased, to form a channel between the two dots;
an end-to-end voltage difference across the nanowire results in an electric field
that propels the electron from source dot to target dot.
We assume that the electron is scattered by phonons, conduction electrons in the wire, and other inelastic scattering sources,
so that it can move from the source dot to the lower-energy target dot without heating up.
We also assume that this scattering can be treated in a Drud\'e-like picture,
implicitly resting on a Boltzmann equation for populations in momentum states.
\begin{figure*}
    \includegraphics[width=0.95\textwidth]{cartoon.pdf}
    \caption{\textbf{Sources of decoherence.}
    \textbf{(a) Spin-orbit coupling:} as the electron scatters, the effective field due to spin-orbit coupling changes randomly.
    We find that this is the dominant source of decoherence.
    \textbf{(b) Spatially-varying $g$-factor:} the effective $g$-factor varies along the channel; scattering processes mean the electron spends an unpredictable amount of time in different effective $g$-factors.
    \textbf{(c) Diabatic valley transitions:} The valley splitting has a large random component, due to (e.g.) alloy disorder at the interface, so sometimes it is small. At points where it is small, the electron can diabatically transition to an excited state, which has a slightly different $g$ factor.
    \textbf{(d) Spin-valley hotspots:} When the valley splitting is on resonance with the Zeeman field, spin-valley coupling can flip both spin and valley; later spin-independent valley relaxation (e.g. due to phonons) brings the electron back to the valley ground state, now in a different spin state.
    \textbf{(e) Nuclear spin impurities:} As the electron passes near the rare spin-1/2 $\si{29}$, hyperfine interaction with the $\si{29}$ nuclear spin can change the electron spin.
    } 
    \label{fig:decoherence-sketches}
\end{figure*}

The shuttling process is necessarily momentum-incoherent, but we expect it to be spin-coherent.
Momentum scattering does not directly threaten spin coherence because many sources of scattering (e.g. phonons or oxygen impurities \cite{miMagnetotransportStudiesMobility2015}) do not directly couple to the spin.
But other effects do threaten spin coherence (Fig.~\ref{fig:decoherence-sketches}).
These include coupling of the spin to momentum via spin-orbit coupling such that the momentum will gradually thermalize the spin
\cite{perelSPINORIENTATIONELECTRONS,dyakonovm.i.SPINRELAXATIONCONDUCTION1972,g.e.pikusSpinRelaxationOptical1984};
the $g$-factor variation in space through the channel;
diabatic transitions between the electron's two valley states give rise to a fluctuating $g$-factor;
and finally nuclear spin impurities can flip or dephase the spin.
We estimate the effect of these noise sources on transport (but not tunneling between dot and channel). 
We find that nuclear spin impurities and valley effects are made negligible by motional narrowing.
Spin-orbit coupling to the momentum, therefore, is likely to be the dominant effect; if the spin-orbit coupling coefficients are comparable to those measured \cite{wilamowskiSpinRelaxationGfactor2003c} and simulated \cite{nestoklonElectricFieldEffect2008} in Si/SiGe heterostructures, it will result in fidelities $\approx \depends{0.996}$ for a \depends{100 $\mu$m} shuttling distance.

The paper is organized as follows.
In Sec.~\ref{s:device} we describe the proposed device and our model of the moving electron's dynamics,
and in Sec.~\ref{s:results} we summarize our results.
In Sec.~\ref{s:soc} we describe the effect of spin-orbit coupling,
in Sec.~\ref{s:valley} we describe valley effects,
in Sec.~\ref{s:spatial-g} we consider spatial variation in the $g$-factor,
and in Sec.~\ref{s:nucl} we describe nuclear spin impurities.
We conclude in Sec.~\ref{s:conclusion} by summarizing our assumptions and results,
outlining experiments which test those assumptions,
and describing some potential applications of the device.

\section{Device}\label{s:device}

\begin{table}[t]
  \centering
  \begin{tabular}{|rcc|}
    \hline
    Si thickness & $\qquad b\qquad$ & 10 nm    \\
    Channel width & $w$ & 100 nm  \\
    Channel length & $L$ & 100 $\mu$m \\
    \hline
    Temperature & $T$ & 30 mK \\
    Electric field in channel & $E$ &1 V/cm\\
    Zeeman field & $B$ & 0.5 T \\
    \hline
    Spin-orbit coupling (Rashba) &$\alpha $\quad & 0.2 peV $\cdot$ cm\\
   (Dresselhaus) &$\beta$& 1 peV $\cdot$ cm\\
    Intervalley $g$-factor variation & $\delta g/g$ &  $10^{-4}$\\
    Mobility & $\mu$ & 100,000 cm${}^2$/V$\cdot$ s \\
    $\si{29}$ concentration &&800 ppm \\
    \hline
  \end{tabular}
  \caption{\textbf{Model device parameters}: except where specified, we take the device to have these parameters.
    For the spin-orbit coupling, see 
    App.~\ref{app:soc:value};
    for the intervalley $g$-factor variation see \cite{ferdousValleyDependentAnisotropic2018}.
}
  \label{tab:device-params}
\end{table}

We consider a layered $\ce{Si/SiGe}$ heterostructure (Fig.~\ref{fig:summary} top), with electrons confined to a $b = 10\ \mathrm{nm}$ layer of Si isotopically enriched to 800 ppm $\si{29}$.
On top of the device are accumulation gates, barrier gates, and a resistive topgate.
These gates form two quantum dots connected by a long channel, formed by biasing the resistive topgate;
each dot is separated from the channel by a barrier controlled by a barrier gate.
The channel is narrow ($w = 100$ nm),
so orbital excitations in the plane of the quantum well but perpendicular to the wire
are gapped out by $\sim \hbar^2/2mw^2 \approx 20\; \mu$eV or a temperature of $\sim 230$ mK, larger than our model device temperature of 30 mK.
{ (If thermal excitations to the transverse orbital excited state become significant at our channel width, we expect that they can be mitigated by fabricating a narrower channel.)}
We therefore take the channel as effectively one-dimensional, 
Table~\ref{tab:device-params} gives the device parameters we use; Table~\ref{tab:derived-params} gives certain derived quantities.

By running a current through the resistive topgate, one can induce an electric field $E$ along the length of the channel. Here, we find it convenient to consider the electric field in the channel as opposed to potential difference to prevent introducing a spurious length dependence.

Electrons in the silicon forming the quantum dots and the channel have a band structure with six conduction-band minima, or valleys, in the $\pm x,y,z$ directions.
Strain effects raise the energy of the $\pm x,y$ valleys by meV, so they are negligible,
leaving the $z$ valley states.
These $z$ valley states are split by the combined effect of the interface and the confining electric potential of the topgate \cite{zwanenburgSiliconQuantumElectronics2013,burkardSemiconductorSpinQubits2023}.
We therefore consider a one-dimensional effective model for the momentum and valley degrees of freedom
\begin{align}
  H_{\text{mom}} = -\frac {\hbar^2k_x^2}{2m_t} + \frac 1 2 \Delta(x)[\tau^x \cos \phi(x)  + \tau^y \sin \phi(x)]\;,
\end{align}
where $m_t = 0.19 m_e$ is the transverse effective mass of the electron,
$\tau^{x,y}$ are Pauli matrices in valley space,
and $\Delta(x)$ and $\phi$ are the the position-dependent valley splitting and phase \cite{losertPracticalStrategiesEnhancing2023a}.

\begin{table}[t]
  \centering
  \depends{
  \begin{tabular}{|rcc|}
    \hline
    Momentum relaxation time & \quad$\tau_k = \mu m_t/e$\quad & 11 ps\\
    Larmor freq.~(angular) & $\Omega = g\mu_B B / \hbar$ & 88 GHz\\
    Larmor freq.~(rotational) & $g\mu_B B / (2\pi\hbar)$ & 14 GHz\\
    Electron Larmor period & $2\pi \hbar/(g\mu_B B_z)$ & 71 ps\\
    Average electron speed & $v_0 = \mu E$ & $10^3$ m/s\\
    Average electron momentum & $k_0 = \mu m E / \hbar$ & 1.6 $\mu\mathrm{m}^{-1}$\\
    Variance of electron momentum & $ \expct{\delta k^2} = \frac{mTk_B}{\hbar^2}$ & $(2.5\;\mu\mathrm{m}^{-1})^2$ \\
    Spin-orbit coupling length & $\hbar^2/m_t\beta$ & $40\ \mu$m\\
    Spin-orbit coupling velocity & $\beta/ \hbar $ & $15$ m/s\\
    Time across 100$\mu$m interconnect & $T_{\text{cross}}$ & 100 ns\\
    \hline
  \end{tabular}
  }
  \caption{\textbf{Derived device parameters} (cf Table \ref{tab:device-params}).}
  \label{tab:derived-params}
\end{table}

The electron is strongly scattered,
resulting in a momentum correlation time $\tau_k = \mu m_t/e$, $\mu$ the mobility;
 we take this to be described broadly by the Drud\'e model.
We assume the scattering is primarily inelastic.
Inelastic scattering requires scattering mechanisms other than the impurities discussed in \cite{miMagnetotransportStudiesMobility2015,huangUnderstandingDisorderSilicon2023}%
---perhaps excitation of phonons or scattering from conduction electrons in the resistive topgate.
Some degree of inelastic scattering is crucial to our scheme, because the electron starts at high energy at one end of the channel and ends at low energy at the other end of the channel;
the inelastic scattering also justifies our Drud\'e picture of transport.
If impurities are in fact the dominant scattering mechanism in our system,
 the electron will Anderson localize, and
electron transport will be better treated in terms of Mott variable-range hopping between Anderson eigenstates
\cite{andersonAbsenceDiffusionCertain1958,mottConductionNoncrystallineSystems1968,leeDisorderedElectronicSystems1985};
we leave that regime to future work.

The electron is in a Zeeman field $B$; we take $B$ to be oriented along the channel, i.e. the $x$ axis.
This gives a Hamiltonian term
\begin{align}
  H_B = \frac 1 2 \mu_B [g + g'(x) + (\delta g)(\tau^x \cos \phi + \tau^y \sin \phi) ] B \sigma^x\;,
\end{align}
where $g'(x)$ gives the (random) spatial variation of the $g$-factor.
(Although this spatial dependence results from variation in the spin-orbit coupling coefficients, it is convenient to treat it separately from the spin-orbit coupling to the incoherent momentum state.
{ Note, we assume that the magnetic field is in the plane of the well.
In an out-of-plane magnetic field, the effective $g$-factor depends on the shape of the electron wavefunction \cite{ruskovElectronGfactorValley2018a}, which would complicate our analysis.})
$\delta g$ encodes the valley dependence of the $g$-factor, which we discuss in Sec.~\ref{s:valley}; in that section we also briefly discuss the effect of spatially-varying $g$ factors, as well as valley hot spots.

The Si/SiGe interface gives a spin-orbit coupling
\begin{equation}\label{eq:Hsoc}
  H_{\text{SOC}} = k_x(\alpha \sigma^y + \beta \sigma^x)\;.
\end{equation}
(In principle it also leads to terms $k_y[\alpha\sigma^x - \beta\sigma^y]$,
but because the electron is in a coherent $k_y$ state, these terms will not lead to irreparable loss of fidelity.)
The valley dependence of the spin-orbit coupling does not affect the decoherence,
as we discuss in Sec.~\ref{ss:valley-soc-g}.

{ The longitudinal part of the spin-orbit coupling $\beta \sigma^x$ can be removed by a gauge transformation (cf \cite{aleinerSpinOrbitCouplingEffects2001}) under the assumption that the momentum scattering rate $\tau_k$ is momentum-independent.
In this case, the longitudinal spin-orbit coupling induces a coherent, predictable longitudinal rotation that depends solely on the distance between source and target dot, and not on the details of the electron's momentum history.
But momentum-dependent scattering leads (in the new gauge) to a direct spin-bath coupling or a gauge-induced spin-orbit coupling. 
In that case the gauge transformation, far from reducing statistical uncertainty by decoupling the momentum,
introduces statistical uncertainty in the spin state.
Put another way---if the scattering is momentum dependent, then the correct gauge transformation depends on the bath state, which is not known.
We expect that effects of this nature make the gauge transformation inapplicable,
so we proceed with an ordinary \DP analysis.
In App.~\ref{app:gauge} we discuss the gauge transformation, questions of momentum dependence,
and implications for our estimates.
}

We assume that the momentum degree of freedom relaxes to a Gibbs state
\begin{align}\label{eq:k-gibbs}
  \rho_{\text{momentum}} = Z^{-1} \exp\left[- \frac {\hbar^2} {2m_tk_BT} (k - k_0)^2\right]\;,
\end{align}
where $\quad k_0 = mE\mu/\hbar$ is the drift momentum.
(This is in contrast to \cite{boscoHighfidelitySpinQubit2024}, which treats conveyor-mode shuttling and so assumes that the electron is always in a coherent position state.)
Eq.~\eqref{eq:k-gibbs} can be understood as the maximum-entropy state subject to energy and average-momentum constraints.

\begin{table*}[t]
  \centering
  \depends{
    \input{summary-results-table.tex}
    }
    \caption{\textbf{Summary of contributions to infidelity} and relaxation ($L_1$) and decoherence ($L_2$) lengths.
    Numerical results use the model device parameters of Table~\ref{tab:device-params}.
    Spatial $g$-factor variability and valley-dependent $g$-factor lead only to decoherence; likewise, spin-valley hotspots lead only to relaxation.
    Although nuclear spins cause both relaxation and decoherence, we only estimate decoherence; we expect relaxation to be an even smaller effect. 
    $\Omega = \mu_B B g/\hbar$ is the (angular) Larmor frequency, $\tau_k = \mu m_t / e$ the momentum scattering time,
    and $\zeta$ the distance between points of small valley splitting. 
    }
    \label{tab:summary}
\end{table*} 

We are agnostic as to the details of the scattering process.
Instead of considering those details, we use a phenomenological momentum correlation function.
We take that correlation function to have single-time variance $\expct{(k-k_0)^2} = m_tk_BT/\hbar^2$, in accordance with \eqref{eq:k-gibbs}, and a correlation time $\tau_k = \mu m_t/e$.
We do assume that the scattering process is inelastic, and in fact thermalizes the electron with the scattering source, so the electron's temperature does not increase as it travels down the wire.

We assume that $\ce{{}^{29}Si}$ nuclear spin impurities are uniformly distributed at a density of 800 ppm  \cite{deelmanMetamorphicMaterialsQuantum2016a} throughout the slice of silicon in which the electron travels.
(We ignore the $\ce{{}^{73}Ge}$ nuclear spins.)
We take the nuclear spins to be at infinite temperature.
These nuclear spins interact with the electron via a contact hyperfine interaction,
which we treat in scattering theory.

\section{Summary of results}\label{s:results}

\subsection{Relaxation and decoherence lengths; Kraus operators}
Qubit performance can be characterized by timescales for spin relaxation ($T_1$) and dephasing ($T_2$).
For an interconnect the parameter of interest is not a dwell time but the interconnect length.
We therefore frame our results in terms of lengthscales for spin relaxation ($L_1$) and dephasing ($L_2$):
\begin{align}
  \begin{split}
    \expct{\sigma^x} &\propto e^{-L/L_1} \\
    \expct{\sigma^{y,z}} &\propto e^{-L/L_2} 
  \end{split}
\end{align}
(recall that our quantization axis is the $x$ axis).
We consider decoherence due to spin-orbit coupling, valley-dependent $g$-factor, spin-valley hot spots, spatial $g$-factor variation, and nuclear spin impurities.
Fig.~\ref{fig:decoherence-sketches} shows sketches of these processes;
Table~\ref{tab:summary} shows their $L_1$, and $L_2$, as well as their contributions to the entanglement fidelity (\textit{vide infra}).

We find that both relaxation and decoherence are dominated by the D'yakonov-Perel' effect, i.e. spin-orbit coupling to the incoherent momentum.
The resulting lengths are
\begin{align}\label{eq:soc-L1-L2-summ}
  \begin{split}
    \frac 1 {L_1} &=  \frac{4m_t^2k_BT}{\hbar^4eE} \frac {\alpha^2} {1 + (\mu_B B g \mu m_t/\hbar e)^2}\\
    \frac 1 {L_2}  &= \frac{2m_t^2k_BT}{\hbar^4eE} \left(2\beta^2 +  \frac {\alpha^2} {1 + (\mu_B B g \mu m_t/\hbar e)^2}\right)\;;
  \end{split}
\end{align}
for the model device parameters of Table~\ref{tab:device-params}, these are
\begin{align}
  \begin{split}
    L_1 &\approx 74\ \mathrm{cm} \qquad \textnormal{(spin-relaxation length)}\\
    L_2 &\approx 15\ \mathrm{mm} \qquad\textnormal{(spin-dephasing length)}\;.
  \end{split}
\end{align}
For reasonable parameters the valley-dependent $g$-factor gives an $L_2$ length $\sim$10 cm, longer than the SOC $L_2$ (Sec.~\ref{ss:valley-soc-g}); the effects of valley hot spots (Sec.~\ref{ss:spin-valley}) and nuclear spin impurities (Sec.~\ref{s:nucl}) are much smaller still.

The resulting quantum channel has Kraus  operators
\begin{align}\label{eq:summary-Kraus}
  \begin{split}
    K_1 &= a\sigma ^- \\
    K_2 &= f(bP_\up + P_\dn) \\
    K_3 &= d(b P_\up - P_\dn)\;.
  \end{split}
\end{align}
with
\begin{align}
  \begin{split}
    a &= \sqrt{1 - e^{-L/L_1}} \\
    b &= \sqrt{e^{-L/L_1}} \\
    d &= \sqrt{ \frac 1 2 \left(1 - e^{-L/L_2} \right) }\\
    f &= \sqrt{ \frac 1 2 \left(1 - e^{-L/L_2} \right) }\;.
  \end{split}
\end{align}

\subsection{Entanglement fidelities and spin blockade}
The \textit{entanglement fidelity} \cite{schumacherSendingQuantumEntanglement1996,nielsenEntanglementFidelityQuantum1996} of a quantum channel $\mathcal E$ is the fidelity of the channel, when applied to one half of a maximally entangled state. In our case, write $\ket{\Omega}$ for a singlet state; the entanglement fidelity is
\begin{align}
  F_e(\varepsilon) = \bra{\Omega}\ (I \otimes \mathcal E)\big(\ketbra{\Omega}{\Omega}\big)\ \ket{\Omega}\;;
\end{align}
it is related to the average of the channel fidelity between single-qubit states 
\begin{align}
  F_{\text{Haar}}(\mathcal E) &= \int_{\psi \in \text{Haar} } d\psi\; \Big\langle \psi \Big| \mathcal E\big( \ketbra{\psi}{\psi} \big) \Big| \psi \Big\rangle
\end{align}
by
\begin{align}
    F_{\text{Haar}}(\mathcal E) = \frac{d F_e(\mathcal E) + 1}{d+1}\;,
\end{align}
$d = 2$ the onsite Hilbert space dimension
(see e.g. \cite[Ex.~48]{meleIntroductionHaarMeasure2024})

The entanglement fidelity of the channel with the Kraus operators of Eq.~\eqref{eq:summary-Kraus} in the previous section is
\begin{align}
  \begin{split}
    F_e &= \frac 1 4 \sum_j \Big(\tr K_j\Big)^2 \\
        &= \frac 1 2 \left(1 + e^{-L/L_2}\right) e^{-L/2L_1} + \frac 1 4 \left(1-e^{-L/2L_1}\right)^2\;.
  \end{split}
\end{align}
In the limit of long relaxation and decoherence lengths, $F_e$ reduces to
\begin{align} \label{eqn:apxFe}
  F_e = 1 - \frac L {2L_2} - \frac L {2L_1} + O[(L/L_{1,2})^2]\;.
\end{align}

\section{Spin-orbit coupling}\label{s:soc}

The spin-orbit coupling term \eqref{eq:Hsoc} in the Hamiltonian couples the momentum, which is in an incoherent Gibbs state,
to the spin state we seek to transport;
this results in relaxation and decoherence due to the D'yakonov-Perel' effect \cite{perelSPINORIENTATIONELECTRONS,dyakonovm.i.SPINRELAXATIONCONDUCTION1972,g.e.pikusSpinRelaxationOptical1984}.
The momentum has a scalar average $k_0$ and a fluctuating part $\delta k$, so the Zeeman and spin-orbit coupling parts of the spin Hamiltonian are
\begin{align}
  \begin{split}
    H_{\text{spin}} = \frac 1 2 g \mu_B B \sigma^x + k_0&(\alpha\sigma^y + \beta \sigma^x)\\
    + \delta k&(\alpha\sigma^y + \beta \sigma^x) + \dots
  \end{split}
\end{align}
The first term is coherent, so we assume it can be corrected by a calibrated rotation after the electron reaches the target dot.
The second term, however, results in a combination of spin relaxation and dephasing.
The magnitude of this effect is controlled by the variance of the momentum $\expct{\delta k^2} = m_tT$;
the value of the spin-orbit coupling coefficients $\alpha, \beta$, 
which we expect to be of order $\alpha \sim 0.2\;\pevcm, \beta \sim 1\;\pevcm$ (cf App.~\ref{app:soc:value});
and the momentum correlation time $\tau_k$ given by the mobility.
We model the effect of spin-orbit coupling by treating the spin degree of freedom as our system of interest,
and the momentum degree of freedom---together with all the scattering sources (impurities, phonons, conduction electrons, etc.)---as the bath.

The rate of relaxation and dephasing depend on the distribution of the momentum fluctuations $\delta k$.
We take $k$ to have a single-time distribution given by the Gibbs state \eqref{eq:k-gibbs}, so $\delta k = k - k_0$ is Gaussian
\begin{align}
  p(\delta k) = Z^{-1} \exp\left[- \frac {\hbar^2} {k_BT} \frac {\delta k^2}{2m_t}\right]\;;
\end{align}
the single-time variance is $\expct{ \delta k^2 } = m_tk_BT/\hbar^2$.
For convenience and definiteness, we take the momentum to have exponential correlations
\begin{align}\label{eq:k-corr}
  \expct{ \delta k(t') \delta k(t) } = \expct{\delta k^2}e^{-|t'-t|/\tau_k}
\end{align}
with correlation time given by the momentum relaxation time
\begin{align}\label{eq:k-corr-time}
  \tau_k = \mu m_t/e\;.
\end{align}
We modify the resulting correlation function to satisfy the KMS relations \cite{kuboStatisticalMechanicalTheoryIrreversible1957a,martinTheoryManyParticleSystems1959,haagEquilibriumStatesQuantum1967},
which are a necessary feature of correlation functions of Hamiltonian systems in thermal equilibrium.
(A careful consideration of scattering processes would likely further modify the correlation function \eqref{eq:k-corr},
but most of those effects are captured by the phenomenological correlation time $\tau_k$ of \eqref{eq:k-corr-time}.
As long as the correlations have that timescale,
we do not expect our results to change by more than an $O(1)$ factor.)
We give details of the resulting correlation function, together with the derivation of the Lindblad equation, in App.~\ref{app:soc-lindblad},
and a brief discussion of the KMS relation in App.~\ref{app:kms}

\begin{figure}
  \includegraphics[width=\columnwidth]{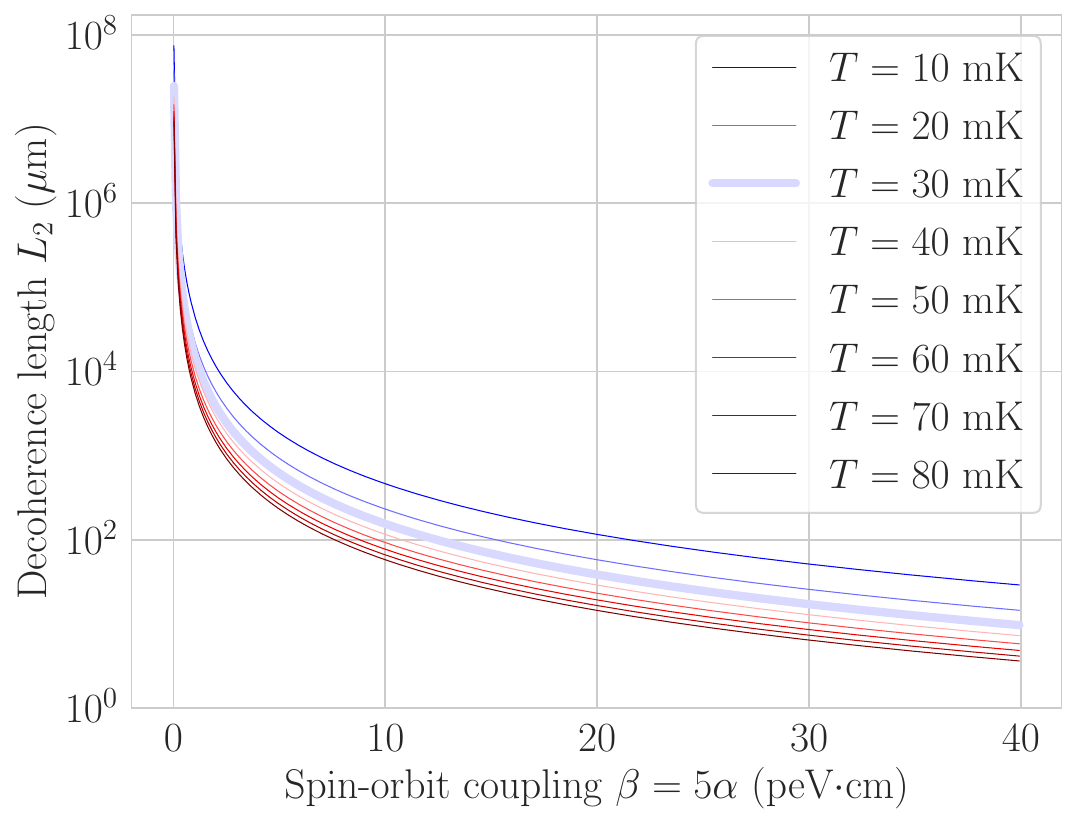}
  \caption{\textbf{Decoherence lengths due to spin-orbit coupling} across electron temperatures. The decoherence length is approximately mobility-independent.
  Because we take the Zeeman field to be along the interconnect, the effective Dresselhaus field is parallel to the quantization axis and decoherence dominates;
  were we to take the Zeeman field perpendicular to the interconnect, the effective Dresselhaus field would still dominate, but would then cause relaxation of approximately the same magnitude.
  }
  \label{fig:soc}
\end{figure}

On timescales long compared with $\tau_k$, the average dynamics can be approximated by Lindblad evolution
\begin{align}
  \label{eq:soc-lindblad}
  \begin{split}
    d_t \rho =\ & \Gamma_{1-}\left( \sigma^- \rho \sigma^+ + \frac 1 2 \{\sigma^+\sigma^-, \rho\}\right)\\
    + & \Gamma_{1+}\left( \sigma^+ \rho \sigma^- + \frac 1 2 \{\sigma^-\sigma^+, \rho\}\right)\\
    + & \Gamma_2\left( \sigma^x \rho \sigma^x - \rho\right)
  \end{split}
\end{align}
with
\begin{align}\label{eq:soc-lindblad-rates}
  \begin{split}
    \Gamma_{1-} &\approx \alpha^2 mk_BT /\hbar^4\frac {4\tau_k}{1 + (\mu_B Bg \tau_k/\hbar)^2} \\
    \Gamma_{1+} &\approx \alpha^2 mk_BT/\hbar^4 \frac {4\tau_k}{1 + (\mu_B Bg \tau_k/\hbar)^2}  e^{-\Omega\hbar/k_BT}\\
    \Gamma_2 &= 2mT\tau_k\beta^2/\hbar^4
  \end{split}
\end{align}
resulting in $T_1$ and $T_2$ rates
\begin{subequations}\label{eq:soc-T12}
  \begin{align}
    \frac 1 {T_1} &\approx 4m_tk_BT \alpha^2\tau_k/\hbar^4 \frac {1}{1 + (\mu_B Bg \tau_k/\hbar)^2} \\
    \frac 1 {T_2} &\approx 2 m_t k_BT \tau_k/\hbar^4 \left(2 \beta^2 + \alpha^2 \frac{1}{1 + (\mu_B Bg \tau_k/\hbar)^2}\right)\;.
  \end{align}
\end{subequations}
(We have ignored $\Gamma_{1+}$, since it is suppressed by a factor $e^{-\Omega \cdot \hbar/k_BT}$ compared to the other rates.)
For momentum scattering time $\tau_k \lesssim \hbar / \mu_B B g$ the Larmor precession time, both $T_1$ and $T_2$ increase as the scattering time decreases, i.e. as the mobility gets lower.
This effect, already noted in \cite{dyakonovm.i.SPINRELAXATIONCONDUCTION1972}, can be understood as motional narrowing in momentum space.
The $T_1$ and $T_2$ of \eqref{eq:soc-T12} lead to relaxation and decoherence lengths
\begin{align}\label{eq:soc-L1-L2}
  \frac 1 {L_1} &= \frac 1 {vT_1} = \frac{4m_t^2k_BT}{\hbar^4eE} \frac {\alpha^2} {1 + (\mu_B B g \mu m_t/\hbar e)^2}\\
  \frac 1 {L_2} &= \frac 1 {vT_2} = \frac{2m_t^2k_BT}{\hbar^4eE} \left(2\beta^2 +  \frac {\alpha^2} {1 + (\mu_B B g \mu m_t/\hbar e)^2}\right)\notag
\end{align}
Fig.~\ref{fig:soc} shows the decoherence length $L_2$ as a function of the spin-orbit coupling coefficients, across temperatures.
(We use the model device values of Table \ref{tab:device-params} for other parameters.)
Because the Dresselhaus spin-orbit coupling $\beta$ is larger than the Rashba $\alpha$ and our magnetic field is parallel to the electron's motion, decoherence dominates.
For the model device parameters of Table \ref{tab:device-params} we find
\begin{align}
  \begin{split}
    L_1 &\approx 0.74\ \mathrm{m} \\
    L_2 &\approx 15\ \mathrm{mm}\;.
  \end{split}
\end{align}

When the momentum scattering time is $\tau_k \lesssim \hbar / (\mu_B B g)$ the Larmor precession time, the SOC relaxation and decoherence lengths are mobility-independent,
because two countervailing effects cancel.
As the mobility decreases the momentum scattering time decreases, leading to the momentum-space motional narrowing discussed above and longer $T_1, T_2$.
But as the mobility decreases, the speed decreases, so the time taken to travel any particular length increases.

The resulting quantum channel has Kraus operators of the same form as Eq.~\eqref{eq:summary-Kraus}.  The entanglement fidelity reduces to Eq.~\eqref{eqn:apxFe}
in the limit of long relaxation and decoherence lengths.

\section{Valley effects}\label{s:valley}
Bulk silicon has six conduction band minima, or valleys: one on each of the $\pm k_x, k_y, k_z$ axes.
In 2D heterostructures the $\pm k_x, k_y$ valleys are gapped out by strain, with gaps of order meV \cite{zwanenburgSiliconQuantumElectronics2013,burkardSemiconductorSpinQubits2023}.
The remaining valleys, on the $\pm k_z$ axis, have a spatially varying energy difference, or ``valley splitting''.
Typically the valley splitting is of order 10-100 $\mu$eV,
comparable to the Zeeman energy,
but \cite{losertStrategiesEnhancingSpinshuttling2024} argues that under plausible assumptions it frequently passes through zero along the length of a device.

The valley degree of freedom can impair device performance by coupling to the electron spin,
together with the magnetic field (a valley-dependent $g$-factor, varying in space)
and electron momentum (valley-dependent spin-orbit coupling, also varying in space).
If the electron is in a mixed valley state, the coupling between valley and spin can dephase the qubit.
If the valley splitting---the energy difference between valley-space energy eigenstates---is near the Zeeman energy,
the device is said to have a \textit{valley hot spot}.
At a valley hot spot, the valley states may hybridize with the spin states,
making the qubit sensitive to charge noise%
\footnote{Typically, the valley states are in slightly different positions (see e.g. \cite{ferdousValleyDependentAnisotropic2018}), so they couple slightly differently to charge noise. When the valley states hybridize with the spin states, then, the spin states also couple slightly differently to charge noise, leading to decoherence.}
and phonon-induced relaxation 
\cite{yangSpinvalleyLifetimesSilicon2013c,huangSpinRelaxationSi2014,tahanRelaxationExcitedSpin2014,haoElectronSpinResonance2014c,cornaElectricallyDrivenElectron2018a,hollmannLargeTunableValley2020a,penthornDirectMeasurementElectron2020,zhangGiantAnisotropySpin2020,hosseinkhaniRelaxationSingleelectronSpin2021,jockSiliconSingletTriplet2022a}.
We argue that the electron is moving too quickly for any of these effects to matter:
motional narrowing makes the effect of valley-dependent spin-orbit coupling and $g$-factor negligible,
and the electron spends too little time near valley hot spots to appreciably relax.

We do not consider valley-driven decoherence after the electron enters the target dot.
If the electron enters the dot's valley-excited state, it may remain there for milliseconds \cite{penthornDirectMeasurementElectron2020}, much longer than relevant gate times;
during that period, the valley $g$-factor difference will give nontrivial decoherence. 

Although the dot entry process requires further study, 
we expect that the probability for the electron to enter the dot valley-excited state will be thermally suppressed.
We assume that the dot entry process will be inelastic, in the sense that it is driven by a process that exchanges energy with some other system
(e.g phonon emission or scattering from wire electrons).
Because that other system is at a temperature small compared to the likely target-dot valley splitting $\Delta_{\text{target}} \sim 100\mu$eV,
the inelastic transition process is much more likely to take the electron to the target dot's valley-ground state than to the valley-excited state.
If the transition process is elastic the electron may end up in the target-dot valley-excited state.
We expect that one could engineer intervalley relaxation, perhaps using a second dot { or a protocol along the lines of that of \onlinecite{odaSuppressingSiValley2024}.
\footnote{ Note, though, that \onlinecite{odaSuppressingSiValley2024} assumes that the velocity---hence the valley degree of freedom and indeed the spin degree of freedom---are coherent, if unknown. In our case the scattering results in an incoherent velocity state, hence an incoherent valley state.}
}

\subsection{Valley-dependent spin-orbit coupling and g-factor}\label{ss:valley-soc-g}
The relevant terms in the Hamiltonian are 
\begin{alignat}{3}\label{eq:valley-sym-terms}
  &H_{\text{spin-valley}} = &\Delta(x) \tau^x\quad & &\text{[valley splitting]} \notag\\
  & &+ \frac 1 2 \hbar\Omega \frac{\delta g}{g}  \tau^x \sigma^x\quad & &\text{[valley-dependent $g$]}\notag \\
  & &+ k \beta'\tau^x\sigma^x \quad& &\text{[valley-spin-orbit]}
\end{alignat}
where $\tau^{x,y,z}$ are valley-space Pauli operators.%
\footnote{
  In fact the valley splitting rotates in the $\tau^{x,y}$ plane, as well as fluctuating in magnitude,
  but this rotation does not affect our estimates.
}
We do not consider valley-dependent Rashba spin-orbit coupling; in effective-mass fits to experimental data, the valley-dependent part of the Rashba coupling is much smaller than the Dresselhaus coupling, which is largely valley-dependent \cite{ferdousValleyDependentAnisotropic2018}.

To estimate the effect of valley-dependent couplings,
we assume that the electron passes diabatically through many places where $\Delta(x)$ is small,
leading to a random valley state.
We assume that small valley splittings occur with density $\zeta^{-1}$
and treat the valley degree of freedom as classical telegraph noise with switching rate $\zeta / v$.
We ignore spatial variation in the coupling constants between valley and spin.%
\footnote{
  This assumption is justified if the correlation length of the coupling constants is not appreciably shorter than $\zeta$, and if the coupling constants and the valley splitting are independent.

  They may not be independent. Consider an interface step.
  \cite{ferdousValleyDependentAnisotropic2018} found the valley-dependent Dresselhaus spin-orbit coupling $\delta \beta$,
  and consequently the valley-dependent $g$ factor $\delta g$, change signs at an interface step.
  But \cite{losertPracticalStrategiesEnhancing2023a} find that under certain conditions the valley splitting has a minimum at an interface step.
  So the effects may cancel: the electron may pass diabatically through this valley-splitting minimuim, leaving the minimum in the valley excited state---but this valley excited state has the same spin-orbit coupling as the previous valley ground state.

  In fact interface steps are unlikely to produce valley minima. \cite{losertPracticalStrategiesEnhancing2023a} finds that if the interface is more than a monolayer thick, the interface steps have a small effect on the valley splitting, which is dominated by alloy disorder, and (in this particular instance) our independence assumption not broken. Nonetheless our independence assumption is nontrivial, and may require reconsideration in future work.
}

The correlation time of the valley state is
\begin{align}
  \tau_{v} \sim \zeta / v_0\;.
\end{align}
For $\zeta = 100$ nm (cf \cite{volmerMappingValleysplittingConveyormode2023,losertStrategiesEnhancingSpinshuttling2024}) and $v_0 = 1000$ m/s, this is $\tau_v \sim 100\;\mathrm{ps} \gg \tau_k \sim 10\;\mathrm{ps}$ the momentum correlation time,
so the effect of Dresselhaus spin-orbit coupling in dephasing the qubit is controlled by the momentum correlation time, not the valley-space correlation time, even though the Dresselhaus spin-orbit coupling is strongly valley-dependent.
For that reason, we have already treated the Dresselhaus spin-orbit-coupling above in Sec.~\ref{s:soc}.

Using the valley correlation time $\tau_v \sim \zeta / v_0$, then, the dephasing rate is
\begin{align}
  1/T_2 &\sim \left(\Omega  \frac{\delta g}{g} \right)^2 (\zeta / v_0)
\end{align}
For reasonable parameters%
\footnote{
The small-valley-splitting length $\zeta$ comes from experiments  \cite{volmerMappingValleysplittingConveyormode2023} with tightly localized electron wavefunctions.
The effective valley splitting is the result of averaging over the valley disorder in the whole area of the wavefunction,
so passing from the tightly localized wavefunctions of \cite{volmerMappingValleysplittingConveyormode2023} to the extended wavefunctions of our device will change the valley-splitting disorder parameters.
If the valley splitting turns out to be an important source of dephasing, this will require more careful consideration.
As it is, we do not expect the resulting change in parameters to change our conclusion that valley effects are less important than spin-orbit coupling.
}
($\zeta = 100$nm \cite{volmerMappingValleysplittingConveyormode2023}, $\delta g / g = 10^{-4}$ \cite{ferdousValleyDependentAnisotropic2018}) this gives $T_2 \sim 100\ \mu$s or $L_2 \sim T_2v \sim 10$ cm, much longer than the $L_2 \sim 15$ mm due to spin-orbit coupling. 

\subsection{Valley hot spots}\label{ss:spin-valley}

In a static dot the fixed valley splitting $\Delta$ may be close to the Zeeman frequency.
If that is the case, and if there is a spin-valley coupling
\begin{align}
  { \frac 1 2 } \Delta \tau^z +  J \tau^z\sigma^z 
\end{align}
between the valley and spin, the two will hybridize.
The spin is then subject to phonon-driven valley relaxation processes and charge noise.
In dots the coupling has been measured at $J \sim 10$ MHz $\cdot h \sim 40$ neV \cite{caiCoherentSpinValley2023}. 

In a resistive interconnect $\Delta$ varies as it passes through different alloy disorder realizations and (potentially) interface steps.
(\cite{losertPracticalStrategiesEnhancing2023a} gives a helpful introduction to this physics.)
If the electron wavepacket has a length scale $l_e$ and a speed $v_0$,
the disorder changes on a timescale $\sim l_e/v_0$;
if the valley splitting $\Delta$ has a standard deviation $\sigma_\Delta$,
its rate of change is
\begin{align}
 \frac{d\Delta} {dt} \sim \sigma_\Delta v_0 / l_e\;.
\end{align}
The spin and valley can only hybridize if
\begin{align}
  | \Delta - g \mu_B B |  \lesssim J
\end{align}
for a time $t_{hs} \gtrsim J/\hbar$.
This timescale
is 
\begin{align}
    t_{hs} \sim J \left[\frac{d\Delta}{dt}\right]^{-1} \sim \frac{Jl_e}{\sigma_\Delta v_0}\;.
\end{align}
The degree of hybridization, then, is roughly
\begin{align}
  p \sim \lambda t_{hs}\sim \frac{J^2 l_e}{\hbar\sigma_\Delta v_0}\;;
\end{align}
if this is small,
then the electron passes through the valley hot spot too rapidly for spin-valley relaxation or valley-mediated charge noise to affect it.

Assume the electron wavepacket has a length scale $l_e \sim 1\; \mu$m, and the standard deviation of $\Delta$ is $\sigma_\Delta \sim 10\;\mu$eV. Then $t_{hs} \sim 0.7$ ps and $Jt_{hs}/\hbar \sim 7 \cdot 10^{-6} $.
We can heuristically estimate a hot-spot driven relaxation length.
Assume that every time the electron passes through a hot spot, its spin state goes to the ground state with probability $p \sim J t_{hs}/\hbar$.
The probability that any particular location is a hot spot is $p_{hs} \lesssim J / \sigma_\Delta$,
so the linear density of hot spots is $\lambda_{hs} \sim J/\sigma_\Delta l_e$ and the relaxation length is 
\begin{align}
    \frac 1 {L_1} \sim p \lambda_{hs} \sim \frac{J^3}{\sigma_\Delta^2 \hbar v_0}  \sim (1\;\mathrm{m})^{-1}
\end{align}
This is comparable to the spin-orbit-driven relaxation length $L_1 \approx 0.74$ m (for our model device parameters), so---depending on the constant factors and the parameters of the actual device---valley hot spots may in fact control spin relaxation.
But it is  much longer than the spin-orbit-driven decoherence length scale $L_2 \approx 15$ mm,
so hot spots are unlikely to limit device performance.

\section{Spatially-varying $g$-factor}\label{s:spatial-g}

Because the $g$-factor has a contribution from spin-orbit coupling it varies slightly with alloy disorder, interface steps, and other random effects.
As the electron passes through the interconnect, then, it experiences a variable $g$-factor.%
\footnote{
Strictly speaking it will experience a variable $g$-factor tensor: that is, the magnetic field term will undergo small random rotations, as well as small variations in magnitude.
We expect this effect to be of the same magnitude as, or even smaller than, the already-small effect of scalar $g$-factor variation, so we do not treat it in detail.
}
The resulting evolution has both a deterministic component and a variable component.
If the electron's path were deterministic, the resulting total $z$ rotation would be the same shot-to-shot, and could be corrected.
But in fact the electron travels diffusively through the channel,
so the path (i.e. the time spent in each region) varies from shot to shot.%
\footnote{In addition, the variable $g$-factor will lead to small spin-dependent scattering, which may lead to further decoherence.
We expect decoherence due to this effect to be smaller than other sources we consider,
because it is second-order: it involves both the (small) spatial $g$-factor variation and some other source of decoherence, e.g. the spin-orbit coupling of Sec.~\ref{s:soc} or the spatially-varying $g$-factor itself.
}

In App.~\ref{app:variable-g} we give a detailed calculation.
In this section we review the assumptions of that calculation
and give a heuristic derivation of the resulting decoherence length.

{
We assume that the electron's position state can be modeled semiclassically,
and that it follows a directed random walk.
We also assume that the variation of the local $g$-factor in the effective-mass model underlying the semiclassical model is random and has a short correlation length.
We expect that variation to be driven by alloy disorder, which modifies the interface properties that control the spin-orbit contribution to the $g$-factor (cf \cite{ruskovElectronGfactorValley2018a}).
Consequently, the local $g$-factor varies on very short length scales.
An extended wavepacket like we consider will average over more of the local disorder than the wavefunctions an electron in a quantum dot,
but in each case the effective $g$-factor is central-limiting.
We can therefore estimate the variability of the local $g$-factor%
---hence the dynamics of the effective $g$-factor controlling the spin dynamics of the electron in either a dot or an extended wavepacket---%
from dot-to-dot variation.
}

Consider an electron in a coherent wavepacket of some length $l_e$,
and assume it travels down the wire in a directed random walk.
The electron coherence length $l_e$ is a key parameter of the calculation, but%
---because it depends on the details of the scattering---%
it is difficult to estimate. 
The coherence length must be longer than the scattering length $\tau_k v_0 \sim 10$ nm;
we expect it to be longer than typical dot length scales, $l_e \gg 100$ nm.
We take the directed random walk to have diffusion constant $D = k_BT\mu/e \approx (160\;\mathrm{nm})^2/\mathrm{ns}$ given by the Einstein relation,
and drift velocity $v = \mu E \approx 1000$ m/s.

Because the wavepacket has length $l_e$ and width $w$ the channel width, the electron spin evolves under the average Larmor frequency in that area.
If the variation in the local Larmor frequency is short-range correlated%
---e.g. because it comes from  spin-orbit coupling variation due to alloy disorder---%
the effective Larmor frequency, averaged over the $l_e \times w$ area of the wavepacket, has correlation length $l_c= 2l_e$\footnote{The factor of 2 comes from a Gaussian integration; see App.~\ref{app:variable-g}.} and location-to-location or device-to-device variance
\begin{align}
    \sigma_\Omega^2 \sim \sigma_0^2 / (l_e w)\;.
\end{align}
Here $\sigma_0$ is a device property which we can estimate from dot-to-dot $g$-factor variation
\begin{align}\label{eq:g-sigma-0}
    \sigma_0^2 \sim l_{\text{dot}}^2 [\Omega_0 \Delta g/g]^2 \approx (0.44\;\mathrm{GHz\cdot nm})^2
\end{align}
where $l_{\text{dot}} \approx 50$ nm is the length scale of the dots of the experiments we consider \cite{jockSiliconMetaloxidesemiconductorElectron2018,mauneCoherentSinglettripletOscillations2012a},
$\Omega_0 = g \mu_B B / \hbar = 88$ GHz is our typical (angular, not rotational) Larmor frequency,
and we take $\Delta g / g = 10^{-4}$ (see App.~\ref{app:larmor-properties}).

To estimate the decoherence length, let us imagine that the Larmor frequency is uniform within boxes of length $l_c$ the Larmor frequency correlation length, but varies randomly between boxes.
The time $t_\cbox$ the electron spends in a box is on average $l_c / v_0$, but has variance
\begin{align}\label{eq:var-tbox}
    \var t_\cbox \sim D l_c / v_0^3\;.
\end{align}
To see this, note that at a time $t = l_c / v_0$ after it enters the left end of the box,
the electron's average position is at the right end of the box,
but the variance of the position is $\var x(t) = D l_c/v_0$.
Since---whatever its position---it is moving at roughly a speed $v_0$,
this means that the variance in its exit time is $\var t_\cbox = \frac 1 {v_0^2} \var x(t)$, giving \eqref{eq:var-tbox}.
(This implicitly assumes that $l_c \gg D/v$, the length scale characterizing when drift dominates diffusive spread.)

If we work in a rotating frame, so the average Larmor frequency is 0, 
then in the box near position $x$, the electron's spin state undergoes a $\sigma^x$ rotation by an angle
\begin{align}
    \theta_{\cbox} = t_\cbox \Omega_x\;.
\end{align}
(Recall that we have taken $x$ to be our quantization axis.)
The shot-to-shot variance of this angle is
\begin{align}
   \var \theta_{\cbox} = \Omega_x^2 \var t_\cbox \sim \Omega_x^2 D l_c / v_0^3. 
\end{align}
The variance of the rotation over the course of the channel, then, is
\begin{align}
    \var \theta_{\text{channel}} = \sum_{\text{boxes}} \var \theta_\cbox
    \sim  L \sigma_\Omega^2 D / v_0^3
\end{align}
since the variances add and the $\Omega_x^2$ average to $\sigma_\Omega^2$.
From this we can read off a decoherence length
\begin{align}
    L_2 \sim \frac {v_0^3} {\sigma_\Omega^2 D} \approx
    \frac{ el_e w \mu^2 E^3}{k_BT\sigma_0^2}\;. %
\end{align}
The more detailed calculation of App.~\ref{app:variable-g} gives the same answer, including constant factors, at leading order in $v_0/(D l_c)$. 
In our case $v_0/D \sim 26$ nm and $l_c \gg 200$ nm, so this leading-order approximation is reasonable.

\begin{figure}[t]
    \includegraphics[width=0.95\columnwidth]{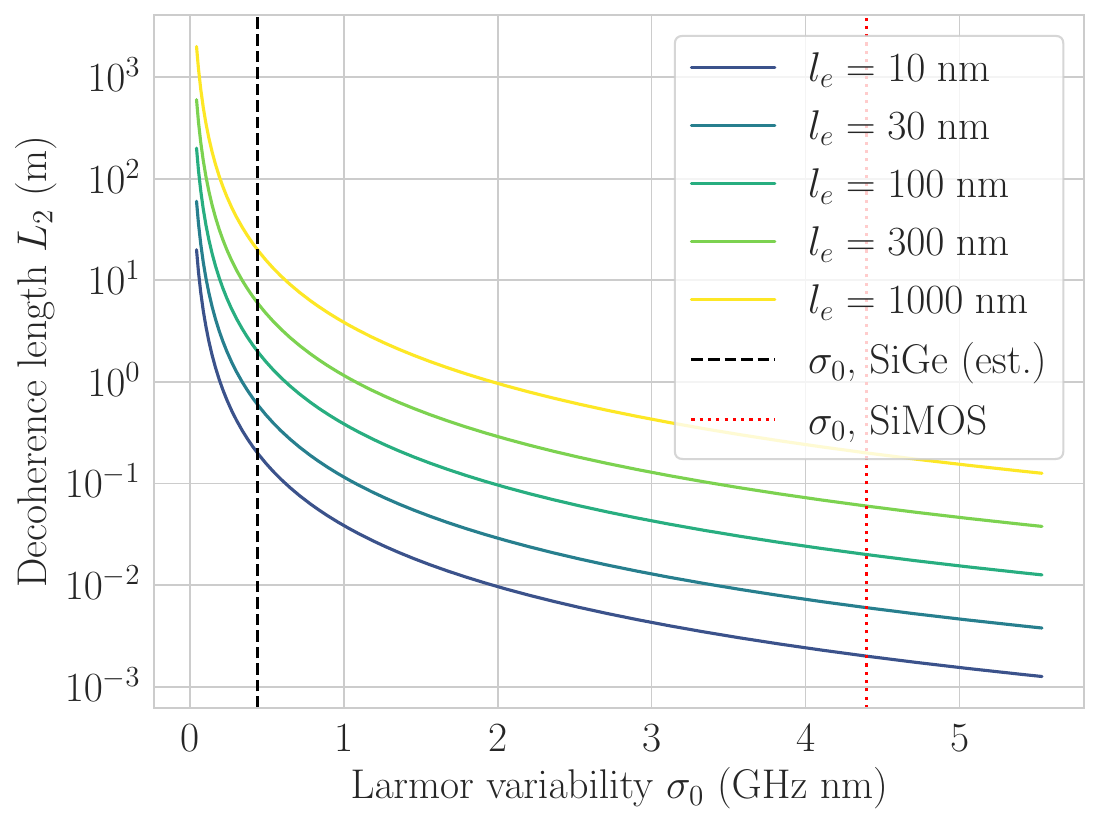}
    \caption{
   \textbf{Decoherence length resulting from spatial $g$-factor variation}, across electron wavepacket lengths.
   Because the Larmor frequency variation is central-limiting, it is best understood as a standard deviation per unit length $\sigma_0$.
   The red dotted line shows $\sigma_0$ corresponding to a dot-to-dot $g$ factor variation $\Delta g / g = 10^{-3}$, consistent with SiMOS measurements \cite{jockSiliconSingletTriplet2022a}; the black dashed line shows a heuristic estimate for SiGe devices, corresponding to $\Delta g / g = 10^{-4}$.
    }
    \label{fig:spatial-g}
\end{figure}

Fig.~\ref{fig:spatial-g} shows this decoherence length as a function of $\sigma_0$ across wavepacket lengths $l_e$.
Under pessimistic assumptions ($l_e = 10$ nm; $\sigma_0 \gtrsim 4$ GHz nm, consistent with SiMOS measurements) this gives a shorter $L_2$ than the spin-orbit coupling contribution we calculated for our model device parameters in Sec.~\ref{s:soc}.
But pessimistic assumptions for the $g$-factor variability are in tension with the model device parameters: if the $g$-factor variability is large, we expect the spin-orbit coupling to be large.
For our model device parameters and $l_e = 100$ nm, the $g$-factor decoherence length is $L_2 \sim 2$ m, much longer than the decoherence length due to spin-orbit coupling $L_2 \approx 15$ mm.

Both this heuristic calculation and the more detailed calculation of App.~\ref{app:variable-g}
make three major assumptions:
that the Larmor frequency variation is short-range correlated;
that the electron's position state can be treated semiclassically, as a wavepacket with a center undergoing a directed random walk;
and that the electron's correlation length is long compared to the mean free path.

The Larmor frequency variation may not be short-range correlated.
It may come from rare, widely-spaced interface steps, or it may come from magnetic field gradients.
Interface steps, like alloy disorder, can change the spin-orbit coupling, hence the $g$-factor.
In that case we expect the calculation to be broadly similar to the calculation we present here.
If the typical step separation $l_{\text{step}}$ is $l_{\text{step}} \ll l_e$, then the calculation here will be unchanged.
If, however, the step separation is $l_{\text{step}} \gg l_e$, then 
the effective $g$-factor will not be central-limiting, so the $\sigma_\Omega^2$ will have a different form, and in the expression \eqref{eq:var-tbox} for $\mathrm{var}\; t_\cbox$ $l_c$ will become $l_\text{step}$.
If the Larmor frequency variation comes from inhomogeneities in the magnetic field, perhaps due to a micromagnet, the tools of App.~\ref{app:variable-g} can still be used, but the calculation will differ substantially.

\section{Nuclear spin impurities}\label{s:nucl}

Although the atoms around the electron are primarily $\ce{^{28}Si}$, which has no nuclear spin,
a few are $\ce{^{29}Si}$, which has nuclear spin $1/2$.
These nuclear spin impurities have a small gyromagnetic ratio, so their spin distribution is not appreciably modified by the Zeeman field,
but they fluctuate on long timescales.

In a slowly-fluctuating Larmor frequency landscape like that caused by nuclear spins
there are two sources of decoherence:
shot-to-shot variations in the amount of time the electron spends in each part of the landscape,
and long-time fluctuations in the landscape itself.
We treated decoherence due to shot-to-shot variations in Sec.~\ref{s:spatial-g}.
(There we focused on $g$-factor variation due to other contributions to disorder in the Larmor frequency, e.g. alloy disorder, but the treatment applies just as well to the nuclear spin contribution.)
In this section we treat decoherence due to fluctuations in the nuclear spins themselves;
we then briefly return to the nuclear spin contribution to the shot-to-shot variation.

The nuclear spins interact with the electron spin via a contact hyperfine interaction \cite{feherElectronSpinResonance1959,burkardSemiconductorSpinQubits2023}
\begin{align}\label{eq:em-hyperfine}
    H_{\text{hyperfine}} = \sum_{\alpha \in \text{impurities}} au\; \delta(x_e - x_n) \bm \sigma_e \cdot \bm \sigma_n 
\end{align}
where $x_e, x_n$ and $\sigma_e, \sigma_n$ are electron and nuclear position operators and Pauli matrices, and
\begin{align}
    au = \frac{2\mu_0}{3} \frac{g\mu_B\gamma_n\eta}{wb}\;,
\end{align}
where $wb$ is the cross-sectional area of the dot (cf Table~\ref{tab:device-params}) and $\eta \sim 100$ is a dimensionless ``bunching ratio''
\cite{shulmanNuclearMagneticResonance1956,wilsonElectronSpinResonance1964,sundforsNuclearMagneticResonance1964,assaliHyperfineInteractionsSilicon2011,philippopoulosFirstprinciplesHyperfineTensors2020}.
(We review the microscopic origin of the effective mass Hamiltonian \eqref{eq:em-hyperfine} in App.~\ref{app:hyperfine}.)
In a dot, this hyperfine interaction leads to Gaussian-in-time dephasing,
because the electron is in contact with the same impurities, in the same spin states, for a long time.
In an interconnect, by contrast, the electron is in contact with each impurity only briefly,
so the dephasing is reduced by motional narrowing.
We consider dephasing due to the term
\begin{align}
    au \delta(x_e - x_n) \sigma^z_e \sigma^z_n\;.
\end{align}

To estimate the magnitude of the dephasing, including motional narrowing,
we take the electron to start in a mixture of wavepackets with some characteristic length $l_e$ and consider the effect of a single nuclear spin.
The electron state is
\begin{align}\label{eq:nucl-wp-mix}
    \rho = \rho_s \otimes \sum_{x,k} p(x) p(k) \ketbra{\psi(x,k)}{\psi(x,k)}
\end{align}
where $\rho_s$ is the electron spin state,
\begin{align}
    \ket{\psi_{xk}} = \int dx'\; e^{-ikx} e^{-(x' - x)^2/4l_e^2} \ket{x'}
\end{align}
is a wavepacket with position $\sim x$ and momentum $\sim k$, $p(x)$ is some position distribution, and
\begin{align}
    p(k) \propto e^{- \frac{\hbar^2}{2mk_BT}(k-k_0)^2}
\end{align}
is the Gibbs distribution for momentum.
Each wavepacket has an interaction matrix element
\begin{align}
    \braket{\psi_{xk}| au\delta(x_e - x_\alpha)| \psi_{xk}} \approx \frac {au}{l_e} 
\end{align}
while the electron is in contact with the impurity.

Ignore, for the moment, momentum relaxation and the electric field.
Then an electron in a wavepacket with momentum $k$ is in contact with the impurity for time
\begin{align}
    t_{\contact} \approx l_e / v = l_em/\hbar k\;.
\end{align}
After averaging over nuclear spin states
the electron's spin state evolves by 
\begin{align}{\label{eq:nuc-dephase-divergent}}
    \rho_s &\mapsto (1 - \Gamma_k) \rho_s + \Gamma_k \sigma^z_e \rho_s \sigma^z_e\;,\notag\\
    \Gamma_k &\approx \left(\frac{au}{\hbar l_e} t_{\contact}\right)^2  = \left(\frac {mau}{\hbar^2k}\right)^2\;.
\end{align}
We give a more careful version of this elementary argument in App.~\ref{app:nucl:elementary-norelax}; we derive the same result in scattering theory in App.~\ref{app:nucl:scattering-theory}.

The dephasing probability $\Gamma_k$ of \eqref{eq:nuc-dephase-divergent} diverges at small $k$.
The divergence indicates that our perturbative approach breaks down for wavepackets with $k \lesssim mau$, and that%
---in the absence of momentum relaxation, and ignoring the effects of the electric field---%
dephasing is dominated by those small-$k$ parts of the momentum distribution.
All of this small-$k$ behavior comes about because at small $k$ the electron is slow, so $t_{\mathrm{contact}}$ is large.

But this scenario, in which the electron remains in contact with a single impurity for a long time, is not realistic.
Because the electron is acted on by the electric field from the resistive topgate and by momentum-relaxation processes,
it is best considered as having a random momentum with correlation time $\tau_k$ the momentum relaxation time.%
\footnote{
Note that in Drud\'e theory the momentum relaxation time is also the time it takes for the electric field to accelerate the electron from a standing start to the steady-state velocity.
}
For times $t \gg \tau_k \approx 10$ ps, then, the electron has average speed $v_0 = \mu E$, so the contact time is roughly $t_\contact \approx l_e / v_0$ and the single-impurity dephasing probability is
\begin{align}
    \Gamma_k \approx (au/\hbar v_0)^2 = \left(\frac{au}{\hbar \mu E}\right)^2
\end{align}
independent of momentum.
(We give a somewhat more careful version of this argument in App.~\ref{app:nucl:elementary-random},
where we assume that the center of each wavepacket undergoes a directed random walk with average speed $v_0 = \mu E$.)
Because the dephasing probability is momentum-independent,
the integration over the momentum degree of freedom $k$ in the initial state \eqref{eq:nucl-wp-mix} is trivial,
and the integration over the position degree of freedom $x$ in \eqref{eq:nucl-wp-mix} is trivial because every wavepacket eventually encounters the impurity.
The dephasing due to this single impurity is therefore
\begin{align}
    \Gamma_{\text{single}} = \left(\frac{au}{\hbar \mu E}\right)^2\;.
\end{align}
The dephasing length is given by the linear density of impurities $\lambda = \nu wb$,
$\nu$ the volumetric density:
\begin{align}\label{eq:nucl-L2}
    \frac 1 {L_2} = \Gamma_{\text{single}} \lambda = \left(\frac{2\mu_0}{3} \frac{g\mu_B\gamma_n\eta}{\hbar \mu E}\right)^2 \frac{\nu}{wb}\;.
\end{align}
At 800 ppm, $\nu \approx 0.04 \;\mathrm{nm}^{-3}$ for
\begin{align}
    \frac 1 {L_2} \approx (1.4\times 10^5\; \mathrm{m})^{-1} \;.
\end{align}

In App.~\ref{app:nucl-dwell-time} we use the results of App.~\ref{app:variable-g} to compute the effect of variation in how long the electron spends near each spin---that is, the disorder average of the shot-to-shot variance, as opposed to the shot-to-shot average of the disorder variance.
We find
\begin{align}
   \frac 1 {L_2}\approx \left(\frac {2\mu_0}{3} \frac{g\mu_B \gamma_n \eta}{\hbar \mu E}\right)^2 \frac \nu {wb} \frac{k_BT}{eEl_e}\;;
\end{align}
For our parameters and $l_e = 100$ nm, this differs from \eqref{eq:nucl-L2} by $T/{eEl_e} \approx 0.26$, so \eqref{eq:nucl-L2} provides an adequate estimate of the effect of nuclear spins.

The dependence $\Gamma \propto (wb)^{-1}$, $wb$ the cross-sectional area of the channel, deserves comment.
This comes about because the hyperfine coupling is $au  \propto |\psi(\bm r_j)|^2 \propto (wb)^{-1} $,
$\psi(\bm r_j)$ the electron wavefunction at the impurity,
and the hyperfine coupling enters the dephasing probability as $(au)^2 \propto (wb)^{-2}$.
As the channel grows wider, that drop in the hyperfine coupling outweighs the increase in the number of impurities the electron contacts.

\section{Conclusion}\label{s:conclusion}

We proposed to use a resistive topgate to connect quantum dots in silicon heterostructures.
When the topgate is biased it creates a 1D channel between the dots;
when a current runs through the topgate, the concomitant electric field drives the electron from source dot to target dot.
We considered dephasing and relaxation of the electron's spin state due to spin-orbit coupling, valley effects, and nuclear spin impurities.
We found that both dephasing and relaxation are dominated by spin-orbit coupling, via the \DP  effect,
leading to characteristic relaxation and dephasing lengths \depends{$L_1 \approx 0.74$ m, $L_2 \approx 15$ mm}.
{ Our estimates could be straightforwardly checked in experiment by performing process tomography of the shuttling process:
by preparing many electrons in the $\sigma^x, \sigma^y$, and $\sigma^z$ eigenstates and measuring their states after shuttling, e.g. by local unitaries and Elzerman readout, one can reconstruct the channel applied by the shuttling process.
This will include not only the incoherent effects we have treated, but also coherent effects like spin-orbit coupling to the average momentum.}

We have focused on transport, not on the one-time dot entry and exit processes.
Tunneling to and from the dot may be slightly spin-selective, due to magnetic field gradients or $g$-factor variation.
In principle this spin-selective tunneling could give rise to an artificial spin-orbit coupling that couples the spin state to charge noise.
In practice, we do not expect this effect to be larger for our devices than for coherent tunneling between dots, especially in the bucket-brigade configuration, which has been demonstrated with high fidelity
\cite{baartSinglespinCCD2016a,flentjeCoherentLongdistanceDisplacement2017,fujitaCoherentShuttleElectronspin2017a,millsShuttlingSingleCharge2019b,mortemousqueEnhancedSpinCoherence2021,yonedaCoherentSpinQubit2021b,noiriShuttlingbasedTwoqubitLogic2022a,zwerverShuttlingElectronSpin2023,vanriggelen-doelmanCoherentSpinQubit2023,smetHighfidelitySinglespinShuttling2024}.

We assume that electron transport is controlled by incoherent, inelastic scattering.
Inelastic scattering is crucial:
the electron moves in a potential landscape created by the resistive topgate, from high potential energy to low potential energy, and that energy must be dissipated by inelastic scattering processes.
We expect that phonon emission, together with scattering between the electron in the channel and the many conduction electrons in the topgate,
provides the necessary dissipation.
We further assume that this scattering can be treated in a Drud\'e-like picture,
implicitly resting on a Boltzmann equation for populations in momentum states.

If transport is controlled instead by static disorder, e.g. the Coulomb disorder treated by \cite{huangUnderstandingDisorderSilicon2023}, our Drud\'e-like picture is not appropriate.
In this case a better picture starts with localized eigenstates, and considers Mott variable-range hopping between those eigenstates.
The spin-orbit calculations of Sec.~\ref{s:soc} no longer apply;
our valley-diabaticity assumptions of Sec.~\ref{s:valley} have to be reexamined;
and nuclear spin impurities become more important, because the electron spends more time in contact with each spin and the motional narrowing of Sec.~\ref{s:nucl} becomes less effective.
Experiments are ultimately needed to determine which regime the device is in.

One may be able to escape this disorder-controlled regime by adding electrons to the channel, which would fill tightly localized low energy bound states.
In this scenario the electron would travel through weakly-localized high-energy states above a Fermi level.
The crossover would be broadly similar to the percolation threshold in 2D materials,
though instead of a percolation transition one would hope merely to weaken the effect of disorder relative to other (inelastic) scattering sources.
Adding electrons to the channel in this way would, however, introduce other sources of decoherence.

The resistive interconnect we propose imposes certain design tradeoffs. 
The interconnect's relaxation and decoherence lengths are controlled by the spin orbit coupling coefficients;
any increase beyond the $\sim 1\;\pevcm$ expected for Si/SiGe wells threatens interconnect fidelities.
So resistive interconnects will require substantial effort in a ``wiggle well'', in which oscillating Ge concentrations in the Si well are
used to control valley splitting and increase spin-orbit coupling \cite{mcjunkinSiGeQuantumWells2022,fengEnhancedValleySplitting2022,woodsSpinorbitEnhancementSi2022,losertPracticalStrategiesEnhancing2023a}.
Resistive interconnects likewise may require more effort in SiMOS or Ge hole devices,
where spin-orbit coupling coefficients are larger than the Si/$\mathrm{Si_xGe_{1-x}}$ devices we consider \cite{ruskovElectronGfactorValley2018a,adelsbergerEnhancedOrbitalMagnetic2022}.
In such systems it may be better to use conveyor-mode shuttling: in conveyor-mode shuttling the electron's position state is coherent, so large spin-orbit coupling protects the electron spin \cite{boscoHighfidelitySpinQubit2024}.

A functioning resistive interconnect will create a number of new architectural possibilities, all requiring further theory work.
In the near term, we look forward to using the resistive topgate for measurement in a solid state Stern Gerlach experiment \cite{barnesQuantumComputationUsing2000,wrobelSternGerlachEffectSpin2006}.
If the topgates form a Y near a micromagnet, the micromagnet will produce a magnetic field gradient
that will divide spin-up and spin-down electrons between the arms of the Y.
Designing such a device will benefit from theoretical modeling to set the device parameters---the micromagnet strength, the electron speed, the width of the channel at the Y, etc.---to maximize the measurement fidelity.

In the longer term, high-fidelity interconnects will enable the genuinely long-range gates required by many error correcting codes.
Our calculations suggest that a resistive interconnect can transport spin states with fidelities { $99.6\%$ over distances of 100 $\mu$m and higher over shorter distances}, depending on spin-orbit coupling and temperature.
If a single qubit in a grid requires 500 nm, this is a range of 20-200 qubits.
But designing a good interconnect architecture for a many-qubit device will require answering many related theory questions, from high-level questions about e.g. routing protocols to low-level device physics questions about e.g. the behavior of the electrons at corners.

\acknowledgements{
We thank
Ben Woods, Merritt Losert, Steve Lyon, Ryan Jock, Jan Krzywda, Mike Winer, Yuxin Wang, Jay Deep Sau, Katharina Laubscher, Michael Swift, Darshana Wickramaratne, and Sasha Efros
for helpful discussions.
  This work was supported in part by ARO grant W911NF-23-1-0242, ARO grant W911NF-23-1-0258, and NSF QLCI grant OMA-2120757. 
 CDW also acknowledges DOE-ASCR Quantum Computing Application Teams program for support under fieldwork proposal number ERKJ347;
 the work was completed while C.D.W. held an NRC Research Associateship award at the United States Naval Research Laboratory.

}
\bibliography{references.bib}

\appendix
\appendixsubsectionformat %

\input{appendix-soc}
\input{appendix-variable-g}
\input{appendix-nucl}

\end{document}

%% file: summary-results-table.tex
 \begin{tabular}{|cccc|}
    \hline
    Source & Relaxation $L_1$ & Decoherence $L_2$ & Infidelity at 100 $\mu$m\\
    \hline
    Spin-orbit coupling & $\frac{4m_t^2k_BT}{\hbar^4 eE} \frac {\alpha^2} {1 + ( \Omega \tau_k)^2} \approx$ 0.74 m\quad & $\frac{2m_t^2k_BT}{\hbar^4 eE} \left(2\beta^2 +  \frac {\alpha^2} {1 + (\Omega\tau_k)^2}\right) \approx$ $1.5 \times 10^{-2}$ m & $3.2 \times 10^{-3}$\\
    Valley-dependent spin-orbit & \multicolumn{3}{c|}{(subsumed by valley-independent SOC )}  \\
    Valley-dependent $g$-factor && $\sim \left(\Omega \delta g / g\right)^2 (\zeta / v^2) \sim$ $1 \times 10^{-1}$ m & $5 \times 10^{-4}$\\
    Spin-valley hot spots & 1 m & & $5 \times 10^{-5}$\\
    Spatial $g$-factor variability & &$2$ m & $2 \times 10^{-5}$ \\
    Nuclear spin impurities && $1 \times 10^{5}$ m & $3 \times 10^{-10}$ \\
    \hline
\end{tabular}

%% file: appendix-soc.tex
\section{Spin-orbit coupling}

\subsection{Value of the spin-orbit coupling coefficient}\label{app:soc:value}

The values of the spin-orbit coupling coefficients $\alpha, \beta$ (for Rashba SOC $H_R = \alpha (\sigma^x k_y - \sigma^y k_x) $ and Dresselhaus-like $H_D = \beta (\sigma^x k_x - \sigma^y k_y)$) depend on many device details, including the electric field, the quantum well thickness, and the interface disorder \cite{pradaSpinOrbitSplittings2011}.
In our model device we take the Rashba and Dresselhaus spin orbit coupling to have values $\alpha = 0.2\; \pevcm$ and $\beta = 1\;\pevcm$ respectively.
This is motivated by conduction electron experiments in Si/SiGe quantum wells at finite electron density \cite{wilamowskiSpinRelaxationGfactor2003c},
$g$-factor anisotropy experiments in Si/SiGe quantum dots \cite{ferdousValleyDependentAnisotropic2018},
and tight binding calculations \cite{pradaSpinOrbitSplittings2011, nestoklonElectricFieldEffect2008,woodsSpinorbitEnhancementSi2022}.

Wilamowski et al. \cite{wilamowskiSpinRelaxationGfactor2003c} report conduction-electron spin resonance (CESR) and spin echo experiments.
(Their device is described in \cite{wilamowskiScreeningBreakdownRoute2001}.)
They apply an external magnetic field at a variable angle to the quantum well;
they find that the dependence of the $g$-factor and the electron $T_1$ and $T_2$ on the magnetic field angle and the filling can all be explained with a single fit parameter, a Rashba spin-orbit coupling coefficient
\begin{align}
\alpha = 0.55\;\mathrm{peV\cdot cm}\;.
\end{align}
They do not consider whether their data can be fit by a Dresselhaus spin-orbit coupling.
Later numerical work \cite{nestoklonElectricFieldEffect2008,pradaSpinOrbitSplittings2011,woodsSpinorbitEnhancementSi2022} finds that these heterostructures are likely to display larger Dresselhaus than Rashba spin-orbit coupling (vide infra).

Ferdous et al. \cite{ferdousValleyDependentAnisotropic2018} measure the $g$-factor anisotropy in a SiGe/Si/SiGe quantum dot,
and estimate spin-orbit couplings via effective mass theory
(and comparisons to atomistic tight-binding models).
They find Dresselhaus and Rashba spin-orbit coupling
\begin{align}
\begin{split}
    \beta &\approx 1\;\mathrm{peV\cdot cm}\\
    { \alpha} &{\approx-0.15\;\mathrm{peV\cdot cm}}
\end{split}
\end{align}
respectively;
{ the Dresselhaus coupling has opposite signs in the dot's two valley states, while the Rashba differs by $\sim 0.03 \;\mathrm{peV\cdot cm}$.}

Tight-binding calculations \cite{nestoklonElectricFieldEffect2008} find%
\footnote{
Note that \cite{nestoklonElectricFieldEffect2008} reverses the notation: they write $\alpha$ for Dresselhaus and $\beta$ for Rashba.
}
\begin{align}\label{eq:soc-tight-binding}
\begin{split}
\alpha &\sim 0.2\; \pevcm \\
\beta &\sim  3.5\;\pevcm\;.
\end{split}
\end{align}
at a transverse electric field $5 \cdot 10^6 \mathrm{V/m}$, for a narrow well of 31 monoatomic planes \cite[Fig.~3]{nestoklonElectricFieldEffect2008}.
The couplings decrease as the well thickness increases; for a larger $20$ nm well, \cite[Fig.~3]{woodsSpinorbitEnhancementSi2022}
find%
\footnote{
    \cite{woodsSpinorbitEnhancementSi2022} goes on to consider so-called ``wiggle wells'', with an oscillating germanium concentration in the well;
    these wiggle wells have higher spin-orbit coupling.
}
\begin{align}
\begin{split}
\alpha &\sim 0.2\; \pevcm \\
\beta &\sim  1\;\pevcm\;.
\end{split}
\end{align}
For wells with clean interfaces the zero-electric-field couplings depend strongly on the number of monolayers in the well, for symmetry reasons
(see \cite{nestoklonElectricFieldEffect2008} for a concise discussion, or \cite{pradaSpinOrbitSplittings2011} for a more pedagogical discussion).
Real wells suffer alloy disorder, which smooths the dependence on the number of monolayers \cite{pradaSpinOrbitSplittings2011}.

\subsection{Lindblad dynamics}\label{app:soc-lindblad}
In general consider
\begin{align}
  H = H_A + H_B + AB\;,
\end{align}
where $H_A$ and $H_B$ act on a system of interest and a bath, respectively, and in the small coupling $AB$ $A$ acts on system and $B$ on bath.
Go to the interaction picture with respect to $H_A$ and $H_B$, treat the coupling perturbatively, and make some standard approximations (Born-Markov and rotating wave).
Then the dynamics of the density matrix for the system of interest is \cite{breuerTheoryOpenQuantum2002}[3.3.1] (cf \cite{schoelkopfQubitsSpectrometersQuantum2003})
\begin{align}\label{eq:lindblad-general}
  \frac d {dt} \rho = \sum_{\omega = 0, \pm \Omega} S_{B}(\omega) \left[A(\omega) \rho A^\dagger(\omega) - \frac 1 2 \left\{A^\dagger(\omega A(\omega), \rho\right\}\right]\;,
\end{align}
where the Lindblad operators are the frequency components of $A$ in an $H_A$ eigenbasis
\begin{align}
  A(\omega) = \sum_{E_{j'} - E_j = \omega} \ketbra{j'}{j'}A\ketbra{j}{j}
\end{align}
and the rates are given by the $B$ correlation function
\begin{align}
  S_B(t) = \tr \left[B(t) B(0) e^{- H_B/k_BT}\right]\;.
\end{align}
Given the approximations that lead to the master equation, this correlation function is the only property of the bath that enters.

In our case the electron spin is the system of interest, and the momentum together with all the scattering sources---impurities, phonons, conduction electrons, etc.---form the bath.
Since $k_0$ is a scalar, the system Hamiltonian is
\begin{align}
  H_A &= \frac {\hbar \Omega} 2 \sigma^x + k_0(\alpha \sigma^y + \beta \sigma^x) \approx \frac {\hbar\Omega} 2 \sigma^x 
\end{align}
(Recall that $\Omega$ is the Larmor frequency $\Omega = \mu_B g B$.)
We ignore the slight perturbation to the spin Hamiltonian due to the coherent part of the spin-orbit coupling.

The correlation function $S_{\delta k}(\omega)$ must satisfy the KMS relation \cite{kuboStatisticalMechanicalTheoryIrreversible1957a,martinTheoryManyParticleSystems1959,haagEquilibriumStatesQuantum1967} (cf App.~\ref{app:kms})
\begin{align}\label{eq:app:kms}
  S_{\delta k}(-\omega) = e^{-\omega/k_BT} S(\omega)\;,
\end{align}
have a correlation time $\tau_k$,
and have a single-time variance
\begin{align}
  mk_BT/\hbar^2 = \expct{\delta k^2} = \int \frac {d\omega}{2\pi} S(\omega)\;.
\end{align}
We take
\begin{align}\label{eq:app:spectral}
  S(\omega) = \frac{mk_BT}{\hbar^2} \frac {4 \tau_k}{1 + (\omega \tau_k)^2} \frac { 1 }{1 + e^{-\omega/k_BT} }\;.
\end{align}
To arrive at this form we take an exponential-in-time correlation decay $\expct{\delta k(t) \delta k(0)} = mTe^{t/\tau_k}$
and convolve it with a filter that enforces the KMS relation \eqref{eq:app:kms};
the KMS relation is more transparent if we write the spectral function in the equivalent form
\begin{align}
  S(\omega) = \frac{mk_BT}{\hbar^2} \frac {2 \tau_k}{1 + (\omega \tau_k)^2} \frac { e^{\omega / 2k_BT }}{\cosh \omega / 2k_BT }\;.
\end{align}
This spectral function gives
\begin{align}
  \begin{split}
    \expct{\delta k^2}
    &= \int_{-\infty}^\infty \frac{d\omega}{2\pi} S(\omega)\\
    &= \begin{cases}
      mk_BT /\hbar^2 & \text{fast decay $T \ll \tau_k^{-1}$} \\
      \frac 1 2 mk_BT/\hbar^2 & \text{slow decay $T \gg \tau_k^{-1}$}
    \end{cases}
  \end{split}
\end{align}
We are in the fast-decay regime:
for our devices
\begin{align}
  \begin{split}
    T \sim 30\;\mathrm{mK} \approx &4\;\mathrm{GHz}\\
    \tau_k^{-1} \approx &100\;\mathrm{GHz}\;.
  \end{split}
\end{align}
Moreover the factor of two difference is smaller than other sources of uncertainty and device-dependence in our calculation,
in particular the spin orbit couplings $\alpha,\beta$.
We therefore do not calculate the interpolating $T$-dependence in the correlation function \eqref{eq:app:spectral}.

The Lindblad master equation, then, is
\begin{align}
  \begin{split}
    d_t \rho =
    \quad \frac 1 {\hbar^2} S_{\delta k}(\Omega) \alpha^2 &\left[\sigma^-\rho\sigma^+ - \frac 1 2 \{\sigma^+\sigma^-, \rho\}\right] \\
    +\frac 1 {\hbar^2} S_{\delta k}(-\Omega) \alpha^2 &\left[\sigma^+\rho\sigma^- - \frac 1 2 \{\sigma^-\sigma^+, \rho\}\right] \\
    +\frac 1 {\hbar^2} S_{\delta k}(0) \beta^2 &\left[\sigma^z\rho\sigma^z - \rho\right] \\
  \end{split}
\end{align}
where $S_{\delta k}$ is the correlation function of \eqref{eq:app:spectral}.
It is convenient to identify rates
\begin{align}
  \begin{split}
    \Gamma_1 = \Gamma_{1-} &= \frac 1 {\hbar} S_{\delta k}(\Omega) \alpha^2 \\
    \Gamma_{1+} &= \frac 1 {\hbar^2} S_{\delta k}(-\Omega) \alpha^2 \\
    \Gamma_2 &= \frac 1 {\hbar^2} S_{\delta k}(0) \beta^2\;;
  \end{split}
\end{align}
these are the rates we quote in Eq.~\ref{eq:soc-lindblad-rates} in the main text.
The $T_1$ and $T_2$ times are
\begin{align}
  \begin{split}
    \frac 1 {T_1} \approx \frac {\alpha^2} {\hbar^2} S(\Omega) &\approx \frac 1 {\hbar^4} \alpha^2mk_BT \frac{ 4\tau_k}{1 + (\Omega \tau_k)^2} \\
    \frac {\alpha^2} {\hbar^2} S(-\Omega) &\approx \frac 1 {\hbar^4}\alpha^2mk_BT \frac{ 4\tau_k}{1 + (\Omega \tau_k)^2} e^{-\Omega/T}\\
    \frac 1 {T_2} = \frac {\beta^2}{\hbar^2} S(0) &= \frac 2 {\hbar^4}mk_BT\beta^2\tau_k
  \end{split}
\end{align}
where we have used that $\hbar \Omega / k_BT \gg 1$ to simplify the KMS parts of \eqref{eq:app:spectral},
and to ignore the contribution of $\dn \to \up$ processes to $T_1$.


\subsection{KMS relation}\label{app:kms}
The KMS relation is a basic feature of Gibbs-state correlation functions.
In the time domain, a correlation function is
\begin{align}
  \expct{A(t) A(0)} = \tr e^{-iHt/\hbar}A e^{iHt/\hbar} A e^{-\beta H}\;,
\end{align}
where for the moment $\beta = 1/k_BT$ is the inverse temperature,
so
\begin{align}
  \begin{split}
    \expct{A(-t) A(0)}
    &= \tr e^{iHt/\hbar}A e^{-iHt/\hbar} A e^{-\beta H}\\
    &= \tr e^{\beta H} e^{-iHt/\hbar} A e^{iHt/\hbar} e^{-\beta H} A e^{-\beta H} \\
    &= \tr e^{-iH(t+i\beta)/\hbar} A e^{iH(t+i\beta /\hbar}) A e^{-\beta H} \\
    &= \expct{A(t+i\beta )A(0)}
  \end{split}
\end{align}
where in the second line we have inserted $1 = e^{-\beta H}e^{\beta H}$ and used the cyclic property of the trace.
(Recall that here $\beta = 1/k_BT$.)
Taking a Fourier transform leads to
\begin{align}
  S(-\omega) = e^{-\omega/k_BT} S(\omega)\;,
\end{align}
the form used in App.~\ref{app:soc-lindblad}.

\subsection{Kraus operators of resulting channel}

In App.~\ref{app:soc-lindblad} we argued that spin-orbit coupling leads to a Lindblad dynamics for the spin
\begin{align}
  \begin{split}
    d_t \rho  &= (\mathcal L_1 + \mathcal L_2)\rho  \\
    \mathcal L_1\rho  &= \Gamma _1\left(\sigma ^-\rho \sigma ^+ - \frac 1 2 \{P_\up, \rho \}\right) \\
    \mathcal L_2\rho  &= \Gamma _2\left(\sigma ^x \rho  \sigma ^x - \rho \right)\;.
  \end{split}
\end{align}
Recall that our quantization axis is $x$, so $\sigma^{\pm} = \sigma^y \pm i \sigma^z$.
As in \ref{app:soc-lindblad} we are ignoring the $\sigma^+$ transition,
because it is suppressed by a factor of $e^{-\Omega/T}$ compared to the other terms.

In this subsection we rehearse a standard calculation for the Kraus operators of the channel given by evolution under this channel for time $t = L/v$.
To find those operators, we compute the action of $e^{(\mathcal L_1 + \mathcal L_2)t}$ on the basis operators $\sigma^x, \sigma^y, P_\up, P_\dn$, and then write Kraus operators that reproduce that action.

\subsubsection{Action of Lindblad evolution on basis operators}
Since
\begin{align}
  \begin{split}
    \mathcal L_1 \sigma^{y,z} &= - \frac 1 2 \Gamma _1 \sigma^{y,z} \\
    \mathcal L_1 P_\up &= -\Gamma_1 (P_\up - P_\dn) \\
    \mathcal L_1 P_\dn &= 0
  \end{split}
\end{align}
and
\begin{align}
  \begin{split}
    \mathcal L_2 \sigma^{y,z} &= -2\Gamma_2\sigma^{y,z} \\
    \mathcal L_2 P_{\up,\dn} &= 0\;,
  \end{split}
\end{align}
the two Lindblad (super)operators commute, and we can treat them separately. Immediately
\begin{align}
  (\mathcal L_1 + \mathcal L_2) \sigma^{y,z} = -\left(\frac 1 2 \Gamma _1 + 2 \Gamma _2\right) \sigma^{y,z}
\end{align}
so
\begin{align}
  e^{(\mathcal L_1 + \mathcal L_2)t}\sigma^{y,z}  = e^{-\left(\frac 1 2 \Gamma _1 + 2 \Gamma _2\right)t} \sigma^{y,z}
\end{align}
from which we can read off
\begin{align}
  \frac 1 {T_2} = \frac 1 2 \Gamma_1 + 2 \Gamma_2\;.
\end{align}
The action on $P_{\up,\dn}$ is less straightforward. Only $\mathcal L_1$ acts nontrivially. Immediately
\begin{align}
  \mathcal L_1^n P_\dn = 0\;,
\end{align}
and one can show
\begin{align}
  \mathcal L_1^n  P_\up = (-\Gamma _1)^n(P_\up - P_\dn)
\end{align}
so
\begin{align}
  \begin{split}
    e^{\mathcal L_1 t} P_\dn &= \sum_n \frac {t^n} {n!} \mathcal L_1^n P_\dn = P_\dn \\
    e^{\mathcal L_1 t} P_\up &= \sum_n \frac {t^n} {n!} \mathcal L_1^n P_\up = e^{-\Gamma _1t} P_\up + (1 - e^{-\Gamma _1t}) P_\dn\;.
  \end{split}
\end{align}
from which we can read off
\begin{align}
  \frac 1 {T_1} = \Gamma_1\;.
\end{align}
Substituting now $t = L/v$ we find that the channel acts by
\begin{align}
  \begin{split}
    \sigma^{y,z} &\mapsto e^{-L/L_2}\sigma^{y,z} \\
    P_\up &\mapsto e^{-L/L_1} P_\up + (1 - e^{-L/L_2}) P_\dn \\
    P_\dn &\mapsto P_\dn
  \end{split}
\end{align}
with
\begin{align}
  \frac 1 {L_1} &= \frac 1 {vT_1} = \frac {\Gamma_1} v \\
  \frac 1 {L_2} &= \frac 1 {vT_2} = \frac {1} v \left(\frac 1 2 \Gamma_1 + 2 \Gamma_2\right)\;.
\end{align}
These are the lengths we quote in the main text (Eq's~\eqref{eq:soc-L1-L2-summ} and \eqref{eq:soc-L1-L2}).

\subsubsection{Channel and Kraus operators}

It is easiest to estimate the channel we seek as the composition of two channels:
\begin{itemize}
\item $\mathcal D_1$ with Kraus operators $a\sigma ^-, bP_\up + cP_\dn$, with the constants $0 \le a,b,c \le 1$ chosen so
  \begin{align}
    \begin{split}
      \mathcal D_1 P_\up &= e^{-\Gamma _1 t} P_\up + (1 - e^{-\Gamma _1t}) P_\dn\\
      \mathcal D_1 P_\dn &= P_\dn \\
      \mathcal D_1 \sigma^{y,z} &= e^{-\frac 1 2 \Gamma_1 t} \sigma ^{y,z}
    \end{split}
  \end{align}
\item $\mathcal D_2$ with Kraus operators $dI, f\sigma ^x$, with the constants chosen so
  \begin{align}
    \begin{split}
      \mathcal D_2 P_{\up,\dn} &= P_{\up,\dn} \\
      \mathcal D_2 \sigma^{y,z} &= e^{-\Gamma _2t} \sigma^{y,z}
    \end{split}
  \end{align}
\end{itemize}
These commute, so
\begin{align}
  \mathcal D_1 \mathcal D_2  = \mathcal D_2 \mathcal D_1
\end{align}
implements the desired channel. One can check that the desired properties come about with
\begin{align}
  \begin{split}
    a^2 &= {1 - e^{-\Gamma _1t}} \\
    b^2 &= {e^{-\Gamma _1 t}} \\
    c^2 &= 1 \\
    d^2 &= { \frac 1 2 \left(1 - e^{-\Gamma _2t} \right) }\\
    f^2 &= { \frac 1 2 \left(1 + e^{-\Gamma _2t} \right) }\;,
  \end{split}
\end{align}
or substituting again $t = L/v$
\begin{align}
  \begin{split}
    a^2 &= {1 - e^{-L/L_1}} \\
    b^2 &= {e^{-L/L_1}} \\
    c^2 &= 1 \\
    d^2 &= { \frac 1 2 \left(1 - e^{-L/L_2} \right) }\\
    f^2 &= { \frac 1 2 \left(1 + e^{-L/L_2} \right) }\;.
  \end{split}
\end{align}

In the first instance the channel we seek has Kraus operators
\begin{align}
  \begin{split}
    K_1' &= fa\sigma ^- \\
    K_2 &= f(bP_\up + c P_\dn) \\
    K_3 &= d\sigma ^x(bP_\up + c P_\dn) \\
         &= (db P_\up - dc P_\dn)\\
    K_4 &= da\sigma^x\sigma ^- \\
         &= - da \sigma ^- \\
         &\sim da \sigma ^-
  \end{split}
\end{align}
where in the last line we have used that Kraus operators are equivalent under sign change. Since $K'_1, K_4$ are scalar multiples of each other and $f^2 + d^2 = 1$ we can write the channel with three Kraus operators
\begin{align}
  \begin{split}
    K_1 &= a\sigma ^- \\
    K_2 &= f(bP_\up + c P_\dn) \\
    K_3 &= d(b P_\up - c P_\dn)\;.
  \end{split}
\end{align}

\subsection{Spin-orbit coupling and gauge transformations}\label{app:gauge}

In the main text we argued that coherent transport was limited by the dephasing induced by Dresselhaus spin-orbit coupling.
But a gauge transformation can remove the spin-momentum coupling; the spin rotation of the electron is then completely determined by the electron's position.
This gauge transformation works under broad but not incontrovertible assumptions---%
in particular, it assumes that the scattering rate is momentum-independent and that the spin-orbit coupling does not depend on any other degrees of freedom.

But the Dresselhaus spin-orbit coupling appears to be strongly valley-dependent \cite{ferdousValleyDependentAnisotropic2018}.
Consequently, to the extent the gauge transformation applies,
we should think about the gauge transformation controlling the spin dynamics \textit{between valley-splitting minima}.
Since the valley, hence the value of the Dresselhaus spin-orbit coupling,
changes randomly at the valley minima, this gives a strong valley-induced decoherence---potentially stronger than the D'yakonov-Perel' decoherence we treated in the main text.

The assumptions underlying the gauge transformation may not hold:
in particular, phonons may give momentum-dependent scattering,
which would break the assumptions of the gauge transformation.
And even if the assumptions do hold,
the resulting valley-induced decoherence can (we expect) be mitigated by rotating the magnetic field.

In this appendix we describe these effects in more detail.

\subsubsection{Removing Dresselhaus spin-orbit coupling by a gauge transformation}

Recall that our electron is described by a Hamiltonian
\begin{align}
    H = \frac {\hbar^2 k^2} {2m} + k(\alpha \sigma^y + \beta \sigma^x) + \frac 1 2 g\mu_B B \sigma^x + V(x,t) \;.
\end{align}
(We have dropped a number of effects here, including scattering. Recall that we take our quantization axis to be $\sigma^x$;
here we write a potential $V(x,t)$ that captures not only the potential slope in the channel, due to the resistive wire,
but also the barriers between channel and dots.)
Under the gauge transformation
\begin{align}\label{eq:gauge-transf}
    \ket{\psi(x)} \to e^{-im ( \alpha \sigma^y + \beta  \sigma^x) x/\hbar^2} \ket{\psi(x)}
\end{align}
the Hamiltonian becomes
\begin{align}\label{eq:app:simple-newgauge-ham}
    H = \frac {\hbar k^2} {2m} +  \frac 1 2 g\mu_B B \sigma^x + V(x,t) \;,
\end{align}
where we have ignored higher-order corrections $\propto \alpha^2, \beta^2, \alpha\beta$, and in the new gauge
\begin{align}
    \sigma^y(x) = e^{im(\alpha\sigma^y + \beta \sigma^x) x /\hbar^2} \sigma^y e^{-im(\alpha \sigma^y + \beta \sigma^x) x /\hbar^2}\;.
\end{align}
Crucially, the Hamiltonian \eqref{eq:app:simple-newgauge-ham} in the new gauge does not couple spin and momentum.%
\footnote{
    This gauge transformation only removes spin-orbit coupling in one-dimensional systems.
    In 2D and higher, the non-commutation of $\sigma^{x,y}$ means that spin-orbit can give nontrivial holonomy%
    ---that is, it can cause a change in spin state when the electron is moved around a closed loop.
    Consequently no gauge transformation can remove it.
    See \cite{aleinerSpinOrbitCouplingEffects2001} for more.
}

Physically, the gauge transformation can be understood as correcting for the effect of the longitudinal (in this case Dresselhaus) spin-orbit coupling.
As the electron moves, the spin-orbit coupling causes its spin state to wind around the $\sigma^x$ axis at a rate $\propto \beta k \propto \beta v$, $v$ the electron speed. 
In a time interval $\delta t$, then,
the electron moves an amount $v \delta x$
and its spin state winds an amount $\propto \beta \delta x$.
This is true even if the momentum (hence speed) changes due to scattering.
Ultimately, the integrated longitudinal-SOC-induced spin rotation is linked not to the details of the momentum history but to the electron's position,
and any dephasing can be eliminated by a simple position-dependent gauge transformation.

The gauge transformation does not materially affect relaxation from (e.g.) a Rashba term perpendicular to the field. 
Under the gauge transformation, a Rashba term becomes
\begin{align}
\alpha k \sigma^y \mapsto \alpha k\ e^{i\beta mx \sigma^x/\hbar^2} \sigma^y e^{-i\beta mx \sigma^x/\hbar^2}\;;
\end{align}
going to the interaction picture with respect to a magnetic field giving Larmor frequency $\Omega$,
this is
\begin{align}
   \alpha \sigma^y(t) = \exp\Big(-i\big[\Omega t - 2\beta mx(t)/\hbar^2\big] \sigma^x\Big) \alpha k \sigma_y\;.
\end{align}
In principle this could change the relaxation processes: if the field (hence $\Omega$) were small compared to $\beta m \dot x(t)\sim \beta m v $, the gauge transformation would be the dominant source of time dependence in $\sigma^y(t)$.
But in our case $2\beta m v \sim 50\; \mathrm{MHz} \ll \Omega \sim 88\;\mathrm{GHz}$, so the $\beta m x$ contribution is negligible and the D'yakonov-Perel' relaxation calculation goes through as described in the main text and the previous part of this appendix.

\subsubsection{Spin orbit coupling gauge transformation and valley-induced decoherence}\label{app:gauge-valley-rotation}

In the geometry we consider, the magnetic field is aligned with the electron's direction of travel.
The gauge transformation then depends on the Dresselhaus spin-orbit coupling. But this can vary widely with the valley degree of freedom, 
from $\beta \to -\beta$,
and the electron randomly switches between valley degrees of freedom at valley-splitting minima.
The valley-splitting minima are spaced by some characteristic length scale $\zeta$.
From one minimum to the next the spin evolves by a unitary
\begin{align}
    e^{i\beta m \zeta \sigma^x/\hbar^2}
\end{align}
with random $\beta$.
The resulting gauge-valley decoherence length is 
\begin{align}\label{eq:gauge-valley}
    \frac 1 {L_2} &\sim [(\beta_+ - \beta_- )m/\hbar^2]^2 \zeta\;,
\end{align}
where $\beta_{\pm}$ are the Dresselhaus SOC in the two valleys.
In Sec.~\ref{s:valley} we take $\zeta \sim 100\; \mathrm{nm}$, based on
\cite{losertPracticalStrategiesEnhancing2023a,volmerMappingValleysplittingConveyormode2023}.
This estimate gives
\begin{align}
    \frac 1 L_2 &\sim (\beta m/\hbar^2)^2 \zeta \sim (15\;\mathrm{mm})^{-1}\;,
\end{align}
(Coincidentally, this is the same as our D'yakonov-Perel' length.) 

The estimate $\zeta \sim 100\;\mathrm{nm}$ sufficed to estimate the magnitude of a sub-leading effect like $g$-factor variation.
But $\zeta$ could be substantially different---perhaps substantially longer---due to the fact that our electrons are not confined in the channel direction.
In the model of \cite{losertPracticalStrategiesEnhancing2023a}, the valley splitting depends on how spread out the electron wavefunction is.
If the wavefunction is more spread out,
it averages over more alloy disorder, so the standard deviation of effective valley splittings is smaller.
(\cite{losertPracticalStrategiesEnhancing2023a} does not emphasize this point, but it is visible in e.g. the $a_\mathrm{dot}$ dependence in Eq.~(E6).)
Consequently the distance between minima will be longer.%
\footnote{ At the same time, because our electrons are quite fast---both the thermal and drift velocities are $\sim 1000$ m/s---the electron may be able to pass diabatically through quite large valley splittings. This will have the countervailing effect of making the distance between relevant valley-splitting minima smaller, because so many more are relevant.}
How much longer depends on the coherence length of the electrons, which in turn depends on the details of the scattering mechanisms, which are not clear.
The coherence length may be as short as $v\tau_k \sim 10$ nm, comparable to the $a_{\mathrm{dot}} \sim 10-20$ nm of \cite{losertPracticalStrategiesEnhancing2023a}, but it may be much longer.
Consequently the gauge-valley decoherence length \eqref{eq:gauge-valley} may be much shorter than the $15$ mm that results from the Dresselhaus \DP effect treated in the main text.

This effect can be mitigated by rotating the magnetic field.
If we take the magnetic field $\boldsymbol{B} = B\hat{\boldsymbol{y}}$, i.e. in the plane of the well but perpendicular to the channel,
then the Dresselhaus SOC gives relaxation---unaffected by the gauge transformation---and Rashba gives decoherence.
The decoherence length is then
\begin{align}
    \frac 1 {L_2} &\sim [(\alpha_+ - \alpha_-) m/\hbar^2]^2 \zeta\;.
\end{align}
But the valley dependence of the Rashba SOC seems to be about two orders of magnitude smaller than that of the Dresselhaus: \cite{ferdousValleyDependentAnisotropic2018} estimate
\begin{align}
\begin{split}
    \alpha_+ &\approx -0.159\;\mathrm{peV \cdot cm} \\
    \alpha_- &\approx -0.126\;\mathrm{peV \cdot cm} \;.
\end{split}
\end{align}
So in this setup the direct (i.e. neglecting relaxation) SOC contribution to the decoherence length is a negligible
\begin{align}
    \frac 1 {L_2} \sim (14\;\mathrm{m})^{-1}
\end{align}
and the relaxation length is the Dresselhaus D'yakonov-Perel' relaxation length, roughly the Dresselhaus $1/L_2\sim 1/(15\;\mathrm{mm})$ of our paper.
(Recall that because the scattering time and the Larmor time are comparable, D'yakonov-Perel' relaxation / decoherence rates do not change much depending on orientation.)

This orientation does come with an additional cost. When the magnetic field is in the well plane but perpendicular to the channel, the Lorentz force modifies how much contact the electron wavefunction has with the interface in a momentum-dependent way;
the SOC parameter and $g$-factor, then, will depend on the momentum.
This effect will introduce a direct spin-momentum coupling.
We expect it to be small.

\subsubsection{Momentum-dependent scattering}
Many effects can complicate the gauge transformation picture of App.~\ref{app:gauge-valley-rotation}.
In particular, if the momentum scattering rate---e.g. due to phonons---is momentum-dependent,
then the gauge transformation leads to a direct spin-bath coupling.
(Momentum-dependent phonon scattering is known to lead to a temperature-dependent renormalization of the gap and other band structure properties \cite[Sections III.A.2, V.B.3, IX.A.1]{giustinoElectronphononInteractionsFirst2017};
these are studied in Allen-Heine perturbation theory.)

Phonons couple to electrons via a term
\begin{align}
    \sum_{k\bm q} g(k,\bm q) c^\dagger_{k+q_x} c_k (a_{q} + a^\dagger_{-\bm q})
\end{align}
where we have introduced momentum-$k$ electron creation and annihilation operators $c^\dagger_k, c_k$ and momentum-$\bm q$ phonon creation and annihilation operators $a^\dagger_{\bm q}, a_{\bm q}$.
We assume that the coupling is spin-independent.
Because the electron is confined to a one-dimensional channel, only momentum along the channel is conserved.
Generically\footnote{
    Some care is required here.
    Our electron is not near a high-symmetry point in the Brillouin zone, but it is on the $\Delta$ line ($k_z$ axis).
    One might therefore expect the matrix element $g(k,\bm q)$ to be symmetric under $k \to -k, q_x \to -q_x$;
    in that case the linear term $g_1$ is forbidden.
    But in fact we expect interface effects---steps and alloy disorder---to break that symmetry;
    they couple to both the electron and the phonons.
    (Recall that germanium has about 2.5 times the atomic mass of silicon.)
    We do not attempt to calculate the magnitude of this effect.
}
one expects the matrix element $g(k,\bm q)$ to expand
\begin{align}
    g(k,\bm q) = g(\bm q) \left( 1 + \frac k {k_1} g_1(\bm q) + \frac 1 2  \left(\frac k {k_1}\right)^2 g_2(\bm q) + \dots\right)
\end{align}
where $k_1$ is an arbitrary reference momentum and $g_1, g_2$ are dimensionless.
Under the gauge transformation \eqref{eq:gauge-transf} the coupling becomes
\begin{align}
\begin{split}
    &\sum_{k\bm q} g(q)\left( 1 + \frac k {k_1} g_1(\bm q) + \frac 1 2  \left(\frac k {k_1}\right)^2 g_2(\bm q) + \dots\right. \\
    &\qquad\qquad\qquad \left. - \frac {\beta m}{k_1} g_1(\bm q) \sigma^x_{\mu\mu'}  - \frac {\beta m}{k_1} g_2(\bm q) k \sigma^x_{\mu\mu'}\right) \\ 
    &\qquad\qquad \times c^\dagger_{k+q_x} c_k (a_{\bm q} + a^\dagger_{-\bm q})
     + O(\beta^2)
 \end{split}
\end{align}
---that is, the first-order term leads to a direct spin-phonon coupling, and the quadratic term leads to a phonon-state-dependent spin-orbit coupling.

We have not attempted to estimate the magnitude of this effect;
it is not even clear that phonons are the dominant scattering mechanism.
(Other scattering mechanisms include conduction electrons in the resistive wire,
disorder in the oxide layer at the device surface,
alloy disorder at the well interface, and more.)

It is possible that these effects are small, and the gauge transformation of \eqref{eq:gauge-transf} gives the appropriate frame in which to work.
We expect that experiments at different magnetic field angles (cf App.~\ref{app:gauge-valley-rotation} above)
could resolve this question, but the interplay of effects is subtle;
we leave that investigation to future work.

%% file: appendix-variable-g.tex
\section{Variable $g$-factor}\label{app:variable-g}

Consider a 1D Hamiltonian
\begin{align}
  H = \sum \frac{p^2}{2m} + \frac 1 2 \hbar \Omega(\hat x) \sigma^z\;.
\end{align}
The Larmor frequency $\Omega(x)$ is random per device and varies in space, hence its dependence on the operator $\hat x$.
Write square brackets
\begin{align}
  [\cdot]
\end{align}
for the average over devices, i.e. realizations of the random Larmor frequencies $\Omega$. 
Work in a rotating frame to null out the average Larmor frequency $[\Omega]$.

Now consider a state
\begin{align}
  \ketbra{\psi _{xk}}{\psi _{xk}} \otimes \rho_s\;,
\end{align}
$\rho_s$ a spin density matrix and $\ket{\psi _{xk}}$ a wavepacket with center $x$, momentum $k$, and length $l_e$, 
At leading order in $\Omega $ the effective dynamics on the spin state is
\begin{align}
  d_t \rho &= - \frac i 2 \underbrace{\braket{\psi _{xk} | \Omega(\hat x) | \psi _{xk} }}_{\equiv \Omega_x} [\sigma^z, \rho_s]\notag \\
           &= - \frac i 2 \Omega _x [\sigma^z, \rho_s]\;.
\end{align}
The effective Larmor frequency is
\begin{align}\label{eq:eff-Omega }
  \Omega _x &\equiv \braket{\psi _{xk} | \Omega(\hat x) | \psi _{xk} } \notag \\
  &= \int dx'\; |\psi _{xk}(x')|^2 \Omega(x')\;.
\end{align}

\subsection{Properties of Larmor frequency disorder}\label{app:larmor-properties}
Assuming the wavepacket is Gaussian with width $l_e$ and the function $\Omega(\hat x)$ is short-range correlated,
the effective Larmor frequency is
\begin{align}
  [\Omega _x] &= 0\notag\\
  [\Omega _x \Omega _{x'}] &= \sigma_\Omega^2 e^{- (x - x')^2 / 2 l_c^2}\;,
\end{align}
for some variance $\sigma_\Omega^2$ and correlation length $l_c = 2 l_e$.
Some later calculations are more transparent if we take $\Omega _x$ to have exponential, rather than Gaussian, correlations
\begin{align}
    [\Omega _x \Omega _{x'}] &= \sigma_\Omega^2 e^{- |x - x'| / l_c}\;.
\end{align}
We expect the difference between exponential and Gaussian correlations to be small compared to the other approximations we are making.

The variance $\sigma_\Omega^2$ is a key parameter.
It is central-limiting, because \eqref{eq:eff-Omega } is an average over a length $l_e$,as well as a width $w$.
So
\begin{align}
  \sigma_\Omega^2 = \frac{\sigma_0^2}{wl_e}\;,
\end{align}
where $\sigma_0^2$ is a real device property.
Since we assume no magnetic field gradients,
the Larmor frequency variation comes from $g$-factor variation;
we can use dot-to-dot variation in the $g$ factor to estimate $\sigma_0$.

We expect $\sigma_0$ to correspond to a dot-to-dot $g$-factor variation $\Delta g / g \sim 10^{-4}$, based on experiments in SiMOS \cite{jockSiliconMetaloxidesemiconductorElectron2018} and SiGe \cite{mauneCoherentSinglettripletOscillations2012a,ferdousValleyDependentAnisotropic2018},
but it may be as large as $10^{-3}$.

\cite{jockSiliconMetaloxidesemiconductorElectron2018} measured $\Delta g / g \sim 10^{-3}$ in SiMOS devices. Since the $g$-factor variation is proportional to the SOC variation and SOC is roughly an order of magnitude larger in SiMOS devices than in the SiGe devices we consider, this result suggests $\Delta g / g \sim 10^{-4}$ for SiGe.
The dots of \cite{jockSiliconMetaloxidesemiconductorElectron2018} are $\sim 50\; \mathrm{nm} \times 50\; \mathrm{nm}$ 
\cite[Suppl.~Fig.~1]{jockSiliconMetaloxidesemiconductorElectron2018}.

\cite{mauneCoherentSinglettripletOscillations2012a}, working in SiGe, do not quote a dot-to-dot $g$-factor difference,
but their results can be used to estimate an upper bound.
They attribute the difference in Larmor frequencies to hyperfine interaction with nuclear spins in their devices, which are made with natural silicon.
They see a standard deviation $\sigma_{HF} = 2.6$ neV, consistent with a priori calculations for the effect of nuclear spins.
Since they work at 30 mT, the Zeeman energy is $\approx 3.5$ neV $\sim 10^3 \sigma_{HF}$, suggesting $\Delta g / g < 10^{-3}$ consistent with our estimate $\Delta g / g \sim 10^{-4}$.
Their electron wavefunctions are somewhat smaller than $50\;\mathrm{nm} \times 50\;\mathrm{nm}$ \cite[Fig.~1]{mauneCoherentSinglettripletOscillations2012a}.

\cite{ferdousValleyDependentAnisotropic2018} measures the difference between the $g$-factors in two valleys, rather than two dots, and find $\Delta g / g \sim 10^{-4}$.
We expect this to be a reasonable proxy for dot-to-dot variation:
the $g$-factor difference is driven by an SOC difference, which is in turn driven by alloy disorder at the interface, and the two valley states see different portions of the interface alloy disorder. 

The result is
\begin{align}
    \sigma_0^2 \sim l_{\text{dot}}^2 [\Omega_0 \Delta g/g]^2 \approx (0.44\;\mathrm{GHz\cdot nm})^2\;.
\end{align}

\subsection{Properties of electron motion}

We assume the electron follows a random walk $\xi_t$ with diffusion coefficient
\begin{align}
  D = T\mu/e \approx (160\; \mathrm{nm})^2/\mathrm{ns}\;.
\end{align}
and drift velocity
\begin{align}
  v_0 = \mu E = 1000\; \mathrm{m/s}\;,
\end{align}
so the probability that it is at position $x$ at time $t$ given that it was at position $0$ at time $0$ is given by the propagator
\begin{align}\label{eq:prop}
  P(\xi_t = x | \xi_0 = 0) = e^{-(x-vt)^2/2Dt}
\end{align}
Drift becomes more important than diffusion on lengthscales larger than
\begin{align}
  l_D = D/v_0 \approx 26\; \mathrm{nm}\;.
\end{align}
Concretely, from \eqref{eq:prop} one can see that once the electron has travelled more than $\text{[a few]} \cdot l_D$, it is unlikely to return to its starting position.
On lengthscales long compared to $l_D$, then, we should think of the electron as primarily following the deterministic drift velocity.

Write
\begin{align}
    \expct{\cdot} 
\end{align}
for the average over $\xi $, i.e. electron paths in single device.

\subsection{Computing decoherence}

Suppose the center of the electron wavepacket follows a random walk $\xi_t$ in the channel.
The electron acquires a total $z$ rotation angle
\begin{align}
  \theta  &= \int_0^\infty dt\; \Omega _{\xi _t} \notag\\
  &= \int_0^\infty dt \int_0^L dx\; \Omega _x \delta(x-\xi _t)\;,
\end{align}
where we take $\Omega  = 0$ in the target dot.
(Recall that $\Omega $ is actually the noise around some nominal value.)
The average over electron paths for a single device is
\begin{align}
    \expct{\theta}
    &= \int_0^\infty dt \int_0^L dx\; \Omega _x \expct{\delta(x-\xi _t)}\notag\\
    &= \int_0^\infty dt \int_0^L dx\; \Omega _x P(\xi _t = x | \xi _0 = 0)\notag \\
    &= \frac 1 {v_0} \int_0^L dx\; \Omega _x \\
  \expct{\theta}^2 &= \frac 1 {v_0^2} \int_0^L dx\;dx'\; \Omega _{x} \Omega _{x'}\;,
\end{align}
where the last step in the $\expct{\theta}$ calculation follows by interchanging integrals and using the diffusion time integral \eqref{eq:diff-time-int}.
The device average is
\begin{align}
  [\expct{\theta}] = \frac 1 {v_0} \int_0^L dx\; [\Omega _x] = 0\;.
\end{align}

The second moment of the rotation angle is
\begin{align}
  \expct{\theta^2} &= \int_0^L dx\; dx'\; \Omega_x \Omega_{x'} \int_0^\infty dt\;dt' \expct{\delta (x - \xi _t) \delta (x' - \xi _{t'})} \notag\\
  &= \int_0^L dx\; dx'\; \Omega_x \Omega_{x'}  G(x,x')\;.
\end{align}
Here
\begin{align}
  G(x,x')
  &= \int_0^\infty dt\;dt' \expct{\delta (x - \xi _t) \delta (x' - \xi _{t'})} \notag\\
  &= \int_0^\infty dt \int_0^\infty d\tau \expct{\delta (x - \xi _t) \delta (x' - \xi _{{t+\tau t'})})}\notag\\
  &\qquad+ (x \leftrightarrow x')\notag\\
  &= \left[\int_0^\infty dt P(\xi _{\tau } = x' - x | \xi _0 = 0) \right]\notag\\
  &\qquad \times \left[\int_0^\infty d\tau P(\xi _t = x'|\xi _0 = 0)\notag \right] + (x \leftrightarrow x')\notag\\
  &= \frac 1 {v_0^2} \left.
    \begin{cases}
      e^{-2|x-x'|} & x' < x \\
      1 & x' > x
    \end{cases}
    \right\} + ( x \leftrightarrow x') \notag \\
  &= \frac 1 {v_0^2} \Big[1 + e^{-2|x-x'|/l_c}\Big]\;.
\end{align}
using the diffusion time integral of App.~\ref{app:diff-time-int}.
Now the second moment is
\begin{align}
  \expct{\theta ^2} &= \frac 1 {v_0^2} \int_0^L dx\; dx'\; \Omega_x \Omega_{x'} \left[1 + e^{-2|x-x'|/l_c}\right] \notag\\
  &= \expct{\theta }^2 + \frac 1 {v_0^2} \int_0^L dx\; dx'\; \Omega_x \Omega_{x'} e^{-2|x-x'|/l_c}\;.
\end{align}
for a variance 
\begin{align}
  \expct{\theta ^2} - \expct{\theta }^2 &= \frac 1 {v_0^2} \int_0^L dx\; dx'\; \Omega_x \Omega_{x'} e^{-2|x-x'|/l_c} \;.
\end{align}
The average \textit{over devices} of the variance \textit{over shots per device} is
\begin{align}
  \Big[\expct{\theta ^2} - \expct{\theta }^2\Big]
  &= \frac 1 {v_0^2} \int_0^L dx\; dx'\; [\Omega_x \Omega_{x'}] e^{-2|x-x'|/l_c}\notag \\
  &\approx L \frac {2\sigma_\Omega^2l_c}{v_0^2} \sqrt{2\pi } e^{4l_e^2/l_D^2} \erfc  (2 l_c / l_D) \quad\text{[Gaussian]}\notag \\
  &\approx L \frac {2\sigma_\Omega^2}{v_0^2} \left[ 1 / l_c + 2/l_D\right]^{-1} \quad\text{[exponential]}
\end{align}
depending on whether the noise has exponential or Gaussian correlations in space.
Assuming $l_D, l_c \ll L$,
the Gaussian case can be treated with the small-$x$ Taylor and large-$x$ asymptotic expansions of erfc
\begin{align}
  \erfc x &\approx
            \begin{cases}
              1 - \frac 2 {\sqrt{\pi }} x & x \ll 1 \\
              \frac 1 {x \sqrt{\pi }} e^{-x^2} & x \gg 1 \;.
            \end{cases}
\end{align}
Since this is somewhat opaque we instead consider exponential correlations,
from which we can read off a decoherence length
\begin{align}
    L_2 = \frac {v_0^2}{\sigma_\Omega^2 l_D} [1 + l_c / 2 l_D]\;.
\end{align}

\section{Diffusion time integral}\label{app:diff-time-int}

We repeatedly encounter the integral
\begin{align}
  &\int_0^\infty dt\; P(\xi _t = x | \xi _0 = 0) \notag\\
  &\qquad = \int_0^\infty dt\; \frac 1 {\sqrt{2\pi Dt}} \exp\left[ - \frac {(x - vt)^2}{2Dt} \right]\;.
\end{align}
To do the integral change variables
\begin{align}
  \begin{split}
    u = \frac v D x \\
    y = \frac {v^2} D t
  \end{split}
\end{align}
for
\begin{align}\label{eq:diff-time-int}
  &\int_0^\infty dt\; P(\xi _t = x | \xi _0 = 0) \notag\\
  &\qquad = \int_0^\infty dt\; \frac 1 {\sqrt{2\pi Dt}} \exp\left[ - \frac {(x - vt)^2}{2Dt} \right] \notag\\
  &\qquad = \frac 1 v \int_0^\infty \frac{dy}{\sqrt{2\pi y}} e^{- (u - y)^2 / 2y } \notag\\
  &\qquad = \frac 1 v \begin{cases} e^{-2|u|} & x < 0 \\ 1 & x > 0\end{cases}\;.
\end{align}

%% file: appendix-nucl.tex
\section{Nuclear spin impurities}\label{app:nucl}

\subsection{Elementary calculation of dephasing by a single impurity, ignoring electric field and momentum scattering}\label{app:nucl:elementary-norelax}

Consider an electron and a single nuclear spin, with the nuclear spin located (for convenience) at $x = 0$.
Take the initial state of the combined system, at some $t_{init} \ll 0$, to be
\begin{align}
    \rho(t_{init}) &= \rho_0\\
    &= \rho_n\rho_s \otimes \int dx\; dk\; p(x) p(k) \ketbra{\psi(x,k)}{\psi(x,k)}  \notag
\end{align}
where $\rho_n, \rho_s$ are the nucleon and electron spin states,
\begin{align}
    \ket{\psi(x,k)} = \int dx'\; e^{-ikx} e^{-(x' - x)^2/4l_e^2} \ket{x'}
\end{align}
is a wavepacket with position $\sim x$, momentum $\sim k$, and width $l_e$,
\begin{align}
    p(k) \propto e^{- \frac{\hbar^2(k-k_0)^2}{2mT}}
\end{align}
is the Gibbs distribution for momentum,
and $p(x)$ is some initial distribution of wavepacket midpoints.
We assume that $p(x)$ is nontrivial only for $x \ll - l_e$, so the wavepackets start well to the left of the impurity.

The system has a Hamiltonian
\begin{align}
    H = \underbrace{\frac {\hbar^2 \hat k^2} {2m} }_{\equiv K}+ \underbrace{au\delta(\hat x) \bm \sigma_e \cdot \bm\sigma_n}_{\equiv V}\;.
\end{align}
(Here we hat the position and momentum operators $\hat x, \hat k$ to distinguish them from the various other position and momentum variables.)

Estimate the electron spin density matrix long after it passes the nuclear spin by treating treating the interaction $V$ as a perturbation, working in the interaction picture, and tracing out the nuclear spin and momentum degrees of freedom.
The full density matrix (of all three degrees of freedom, electron spin, nuclear spin, and electron momentum) is
\begin{align}
\begin{split}
    \rho(t= \infty) = \rho_0 &- i \hbar \int_{-\infty}^\infty dt\;  [V_I(t), \rho_0] \\
    &- \hbar^2\int_{-\infty}^\infty dt\;\int_{-\infty}^t dt'\; [V_I(t),[V_I(t'),\rho_0]]
\end{split}
\end{align}
where we have extended the lower integration limit from $t_{\init}$ to $-\infty$.
Performing the trace, the individual terms are
\begin{align}
\begin{split}
    \tr_{k,n} [V_I(t), \rho_0] &= au \int dx\; dk\; p(x)p(k) \braket{\psi_{xk}| \delta(\hat x(t)) | \psi_{xk} }\\
    &\qquad \times\tr_n [ \bm \sigma_n \cdot \bm \sigma_e \rho_n \rho_s] \\
    &= 0
\end{split}
\end{align}
because $\tr \sigma^j \rho_n = 0$, and
\begin{align}\label{eq:nuclear-double-commutator}
    &\tr [V_I(t)[V_I(t'), \rho_0]]\notag\\
    &\quad = (au)^2 \left[\int dx\; dk\; p(x)p(k) \braket{\psi_{xk} | \delta(\hat x(t)) \delta(\hat x(t')) | \psi_{xk}}  + h.c.\right]\notag\\
    &\qquad\qquad \times (\rho_s - \sigma^z \rho_s \sigma^z)
\end{align}
where for now we are ignoring the relaxation terms $\sigma^x, \sigma^y$ in the interaction.

To compute the overlap $\braket{\psi_{xk} | \delta(\hat x(t)) \delta(\hat x(t')) | \psi_{xk}} $ note that $\ket{\psi_{xk}}$ is in a subspace close to momentum $k$.
Ignoring dispersion, then,
\begin{align}
    \hat x(t) \approx \hat x + (\hbar k/m)t\;,
\end{align}
so
\begin{align}
&\braket{\psi_{xk} | \delta(\hat x(t)) \delta(\hat x(t')) | \psi_{xk}}  \notag\\
&\quad = \int dx'dx'' \psi_{xk}(x')^* \braket{x'| \delta(\hat x + vt) \delta(\hat x + vt' | x''} \psi_{xk}(x'') \notag\\
&\quad = \int dx'dx'' \psi_{xk}(x')^* \braket{x'|x''} \delta(x'' + vt) \delta(x'' + vt')  \notag\\
&\quad = |\psi_{xk}(-vt)|^2\delta(vt-vt') \notag\\
&\quad = |\psi_{xk}(-vt)|^2 v^{-1} \delta(t-t')\;.
\end{align}
where $v = k/m$.
Now
\begin{align}
    \tr [V_I(t)[V_I(t'), \rho_0]] &= \frac{m(au)^2}{k} \delta(t-t') (\rho_s - \sigma^z \rho_s \sigma^z) \notag\\
\end{align}
and the $t = \infty$ electron spin state is
\begin{align}
    &\rho_s(t = \infty) \notag\\
    &\quad = \rho_s(0) - \int dx\; dk\; p(x) p(k)\int_{-\infty}^\infty dt \int_{-\infty}^t dt'\; \delta(t-t')  \notag\\
    &\qquad \qquad\qquad \times |\psi_{xk}(-\hbar k/m \cdot t)|^2\frac{m(au)^2}{\hbar^3 k} (\rho_s - \sigma^z \rho_s \sigma^z) \notag\\
    &\quad = \int dk\;p(k) \Big[ (1 - \Gamma_k)\rho_s(0) + \Gamma_k \rho_s(0) \Big]
\end{align}
The position integral is trivial---physically this came about because every wavepacket eventually encounters, or has encountered, the nuclear spin, and we assume that position and momentum are uncorrelated. We therefore drop it.
We have identified a momentum-dependent dephasing probability 
\begin{align}
    \Gamma_k &= \int_{-\infty}^\infty dt \int_{-\infty}^t dt'\; \delta(t-t') |\psi_{xk}(-\hbar k/m \cdot t)|^2\frac{m(au)^2}{\hbar^3 k}  \notag\\
    &= \left(\frac {mau} k\right)^2\;.
\end{align}
{ (where the $t$ integral is done by change of basis and normalization of $\psi_{xk}$.}
This agrees with the momentum-dependent dephasing probability estimated in the main text, which followed the same logic.
It also agrees with a scattering theory calculation in App.~\ref{app:nucl:scattering-theory}.

\subsection{Elementary calculation of dephasing by a single impurity, with random walk}\label{app:nucl:elementary-random}

In App.~\ref{app:nucl:elementary-norelax} we assumed that a wavepacket $\ket{\psi_{xk}} $ traveled with a constant speed $v = k/m$ in the unperturbed (KE-only) dynamics.
That dynamics gave an interaction-picture position operator
\begin{align}
    \hat x(t) = \hat x + \xi_t\;, \xi = (\hbar k/m) t\;,
\end{align}
where $\xi_t$ was the amount the wavepacket had moved.
This ignored both momentum relaxation and the effect of the electric field.

To account for the effect of momentum scattering and electric field
we now assume that the wavepacket follows a random walk
with average speed $v_0 = \mu E$
and diffusion coefficient given by an Einstein relation $D = T\mu / e$;
for our model device these are
$v_0 = 1\;\mu\mathrm{m/ns}$ and $D \approx (160\;\mathrm{nm})^2/\mathrm{ns}$.
This random walk model is reasonable on timescales $t \gg \tau_k$ the momentum relaxation time.
In this model $\xi_t$, the amount the wavepacket has moved, is a Wiener process with drift.

Although we view the random walk primarily as a phenomenological description of the effect of momentum relaxation and electric field,
it can be justified microscopically.
Translation-invariant scattering sources%
---e.g. phonon emission or scattering from conduction electrons in the resistive topgate---%
will give a Lindblad dynamics
\begin{align}\label{eq:ti-scattering-lindblad}
    d_t \rho = &-i [K,\rho] \\
    &- \int dk\; \gamma_{k'k}\left[c^\dagger_{k'}c_k \rho - \frac 1 2 \{n_k(1-n_{k'}), \rho\} \right] \notag
\end{align}
(or, equivalently, Boltzmann-equation dynamics for the momentum degree of freedom).
$K = k^2/2m$ is the kinetic energy. $\gamma_{k'k}$ is the rate of scattering $k \to k'$;
it depends on matrix elements with the scattering source, the scattering source temperature and density of states, etc.
The Lindblad dynamics \eqref{eq:ti-scattering-lindblad} has many unravelings---that is, there are many different stochastic dynamics that give the same Lindblad equation on averaging.
One convenient unraveling consists of random jumps
\begin{align}
    \ket{\psi_{xk}} \to \ket{\psi_{xk'}}
\end{align}
with rate $\gamma_{k'k}$. In Boltzmann language, this corresponds to sampling from the Boltzmann dynamics of the combined probability distribution $p(x,k)$.
On timescales $t \gg \tau_k$ the relaxation time, these jumps lead to a random walk.


In this random walk picture the relevant matrix element is
\begin{align}
&\braket{\psi_{xk} | \delta(\hat x(t)) \delta(\hat x(t')) | \psi_{xk}}  \notag\\
&\quad = \int dx'dx'' \psi_{xk}(x')^* \braket{x'| \delta(\hat x + \xi_t) \delta(\hat x + \xi_{t'} | x''} \psi_{xk}(x'') \notag\\
&\quad = \int dx'dx'' \psi_{xk}(x')^* \braket{x'|x''} \delta(x'' + \xi_t) \delta(x'' + \xi_{t'})  \notag\\
&\quad = |\psi_{xk}(-\xi_t)|^2 \delta(\xi_t - \xi_{t'})
\end{align}
(compare \eqref{eq:nuclear-double-commutator})
and so---for a particular random walk $\xi$---the momentum-dependent dephasing probability is 
\begin{align}
    \Gamma_k &= (au/\hbar)^2\int_{-\infty}^\infty dt\int_{-\infty}^t dt'\;  |\psi_{xk}(-\xi_t)^2| \delta(\xi_{t'}-\xi_{t}) \;.
\end{align}
But we also need to average over random paths $\xi$.
The path average of the integrand can be re-written
\begin{align}
     &\expct{ |\psi_{xk}(-\xi_t)|^2 \delta(\xi_t - \xi_{t'}) } \notag \\
     &\quad = \int dx'\; |\psi_{xk}(x')^2|\cdot P(\xi_t' = x' | \xi_{t_{init}} = 0) \notag\\
     &\qquad\qquad\times  P(\xi_{t} = x' | \xi_{t'} = x')
\end{align}
{(note that we are abusing notation by writing $\expct{\cdot}$ for the average over paths).}
In general, the the (time- and space-translation invariant) propagator of a Wiener process is
\begin{align}
    P(\xi_{\tau} = y | \xi_{0} = 0) = \frac 1 {\sqrt{2\pi Dt}} \exp\left[-\frac{(y - v\tau)^2}{2D\tau}\right]\;,
\end{align}
so
\begin{align}
    P(\xi_t' = x' | \xi_{t_{init}} = 0)  
    &= \frac 1 {\sqrt{2\pi D (t' - t_{init}) }} e^{x'^2/2D(t' - t_{init})} \notag\\
    P(\xi_{t} = x' | \xi_{t'} = x')
    &= \frac 1 {\sqrt{2\pi D(t-t')}} e^{-v^2(t-t')/2D}
\end{align}
The latter is independent of $x'$---ultimately this is because the random walk is translation-invariant.
So the matrix element is
\begin{align}
     &\expct{ |\psi_{xk}(-\xi_t)|^2 \delta(\xi_t - \xi_{t'}) } \notag \\
     &\quad = \frac 1 {\sqrt{2\pi D(t-t')}} e^{-v_0(t-t')/2D} \\
     &\qquad\quad\times\int dx'\; |\psi_{xk}(x')^2|
     \frac 1 {\sqrt{2\pi D (t' - t_{\rm init}) }} e^{x'^2/2D(t' - t_{\mathrm{init}})} 
\end{align}
The $x'$ integral is easily estimated by noting that $t_{\rm init}$ is large,
so most of the time $|\psi_{xk}(x')|^2$ is narrowly spread around $x' = x$
and
\begin{align}
     &\expct{ |\psi_{xk}(-\xi_t)|^2 \delta(\xi_t - \xi_{t'}) }  \\
     &\quad \approx \frac 1 {\sqrt{2\pi D(t-t')}} e^{-v_0^2(t-t')/2D} \times P(\xi_{t'} = x | \xi_{t_{\rm init}} = 0)  \notag
\end{align}
Now
\begin{align}
\Gamma_k &= (au/\hbar)^2 \int_{t_{init}}^\infty dt\; P(\xi_{t'} = x | \xi_{t_{init}} = 0)  \notag\\
&\qquad\qquad\times\int_0^\infty ds\;  \frac 1 {\sqrt{2\pi Ds}} e^{-v_0^2s/2D} \;.
\end{align}
The $s$ integral in the second line is easily seen to be
\begin{align}
    \int_0^\infty ds\;  \frac 1 {\sqrt{2\pi Ds}} e^{-v_0^2s/2D}  = \frac 1 {v_0}
\end{align}
(after a change of variables, one recognizes $\Gamma(1/2)$).
The $t$ integral is the diffusion time integral of App.~\ref{app:diff-time-int}
\begin{align}
    &\int_{t_{init}}^\infty dt\; P(\xi_{t'} = x | \xi_{t_{init}} = 0) \notag\\ 
    &\quad= \int_0^\infty d\tau \frac 1 {\sqrt{D\tau\cdot 2\pi}} \exp\left[-\frac{(x - v_0t)^2}{2Dt}\right]\notag\\
    &\quad= \frac 1 {v_0}\;.
\end{align}
Combining the integrals gives the expected
\begin{align}
    \Gamma_k = \left(\frac {au}{\hbar v_0}\right)^2 = \left(\frac {au}{\hbar \mu E}\right)^2
\end{align}

\subsection{Scattering theory calculation of dephasing by a single impurity}\label{app:nucl:scattering-theory}

Write $\ket k$ for an incoming wavepacket at momentum $k$;
$\ket{L_k}, \ket{R_k}$ are outgoing wavepackets (leftward and rightward) at momentum $k$;
and $\hat R^\sigma _k, \hat T^\sigma _k$ are the action on the spin degree of freedom corresponding to reflection and transmission respectively.
$\sigma  = \pm 1$ labels the nuclear spin.
In general the outgoing momenta do not have to match the incoming momenta.
But the scattering is elastic because we consider dephasing,
{ which does not change the electron energy}.

We again treat the hyperfine interaction as a perturbation.
Consider an eigenstate of the unperturbed Hamiltonian (KE and Zeeman field) 
\begin{align}
    \ket \phi = \ket{kz}\;,
\end{align}
where $z = \pm 1$ is an electron spin eigenvalue.
This state has energy
\begin{align}
  E = \frac {\hbar^2 k^2}{2m} - \frac 1 2 \hbar\Omega z\;,
\end{align}
so the Green's function is
\begin{align}
  \begin{split}
    &\left[H_0 - E - i\epsilon \right]^{-1}\\
    &\qquad= \int dk' \frac {2m/\hbar^2}{k'^2 - k^2 -  (m\Omega/\hbar)(\sigma_e^z -z) - i\epsilon } \ketbra{k'}{k'}
  \end{split}
\end{align}
and the first Born approximation is
\begin{align}
  \begin{split}
  \ket \psi &= \ket k + \int dk' \frac {2m/\hbar^2}{k'^2 - k^2 - (m\Omega/\hbar)(\sigma_e^z -z) - i\epsilon } \ketbra{k'}{k'} \\
  &\qquad\qquad \times -au\sigma^z_n\sigma^z_e \ketbra{r = 0}{r = 0}kz\rangle\;.
  \end{split}
\end{align}
Since $\braket{r = 0|k} = 1$, this is
\begin{align}
  \begin{split}
    \ket \psi &= \ket k - \int dk' \frac {2mu a z/\hbar^2}{k'^2 - k^2 - (m\Omega/\hbar)(\sigma_e ^z -z) - i\epsilon } \ket{k'z} \\
    &= \ket k - \int dk' \frac {2mu az/\hbar^2}{k'^2 - k^2 - i\epsilon } \ket{k'z}
  \end{split}
\end{align}
In real space this is
\begin{align}
\ket \psi = \int dx \left[ e^{ikx} + i\frac{mau z}{\hbar^2k} e^{-ik|x|}\right] \ket{xz}
\end{align}
from which we can read off reflection and transmission matrices
\begin{align}
  \label{eq:T2-unrenorm}
  \begin{split}
    \hat R_k &= i\sigma_n \frac {m au} {\hbar^2 k} z\\
    \hat T_k &= 1 - i\sigma_n \frac{m au} {\hbar^2 k} z\;.
  \end{split}
\end{align}
The scattering rate $mu a / k$ diverges as $k \to 0$.

The scattering process takes
\begin{align}
\begin{split}
  \ketbra{k}{k} \otimes \rho  &\mapsto
  \sum_\sigma  \Big[\ketbra{L_k}{L_k} \otimes \hat R_k^\sigma  \rho  \hat R_k^{\sigma \dagger}\\
  &\qquad\qquad\qquad+ \ketbra{R_k}{R_k} \otimes \hat T_k^\sigma  \rho  \hat T_k^{\sigma \dagger}\Big]\\
  &\qquad+ \text{[L-R cross terms].}
\end{split}
\end{align}
We assume that soon after scattering from the nuclear spin, the electron scatters again from non-magnetic impurities, from conduction electrons in the wire on top of the channel, or from other scattering sources.
That scattering renders the momentum degree of freedom incoherent, so the action on the spin degree of freedom is given by the partial trace,
which can be read off using the fact that all the $\ket{L_k}, \ket{R_k}$ are orthonormal.
The partial trace---the effective action on the electron spin degree of freedom---is therefore
\begin{align}
  \rho  &\mapsto \int dk\; p_k \Bigg[
  \hat R_k \rho \hat R_k^{\dagger} + \hat T_k  \rho  \hat T_k^{ \dagger}\Bigg]\;.\notag \\
\end{align}
Averaging out the nuclear spin and recalling that $z$ is the eigenvalue of $\sigma^z_e$ gives
\begin{align}
\begin{split}
 \rho  &\mapsto \int dk\; p_k \Bigg[(1 - \Gamma_k) \rho + \Gamma_k \sigma^z \rho \sigma^z_e\Bigg]\;,\\
  &\Gamma_k = \left(\frac{mau}{\hbar^2 k}\right)^2\;.
\end{split} 
\end{align}

\subsection{Hyperfine interaction}\label{app:hyperfine}
\subsubsection{Dipole-Dipole interaction with the nucleus}
A magnetic impurity with nuclear spin $\bm I$ (in our case $I = 1/2$) gives rise to a magnetic field
\begin{align}
  \label{eq:B-full}
  B(\bm r) = \frac {\mu _0}{4\pi} \left[\frac {3\hat{\bm r}(\hat{\bm r}\cdot \bm m - \bm m) } {r^3} + \frac {8\pi} 3 \bm m \delta ^3(\bm r)\right]
\end{align}
with $\bm m = \gamma _n I$, $\gamma _n$ the gyromagnetic ratio of the nucleus, and the nucleus at $\bm r = 0$.
Experiments of Feher \cite{feherElectronSpinResonance1959} show that the $\delta $-function part dominates, so we can take this to be
\begin{align}\label{eq:B-contact}
  B(\bm r) = \frac {2\mu _0}{3} \gamma _n \bm I\; \delta ^3(\bm r)\;.
\end{align}
 \cite{haleShallowDonorElectrons1969} gives
\begin{align}
  \gamma _n &= -8.458\; \mathrm{MHz/T} \cdot 2\pi\hbar\notag\\
  &\approx -35\ \mathrm{neV/T}
\end{align}
for the $\ce{^{29}Si}$ gyromagnetic ratio. 

The energy of an electron in this field is
\begin{align}
  \begin{split}
    E &= \int d^3 \bm r |\psi (\bm r)|^2 \mu _B z_e B_z(\bm r) \\
    &= \frac{ 2\mu _0}{3} g\mu _B\gamma _n (z_e  \sigma )|\psi (\bm r)|^2\label{eq:hyperfine-fundamental}
  \end{split}
\end{align}
(assuming, as elsewhere in this appendix, that the nucleus is in a $\pm z$ spin eigenstate $\sigma $).

The question, now, is the wavefunction $|\psi (\bm r)|^2$\;.

\subsubsection{Effective mass theory: wavefunction on scales $r > a$}\label{app:effective-mass}

To predict the wavefunction at the level of unit cells---that is, with a ``resolution''  $R \gtrsim a$ the lattice constant---use effective mass theory:
solve
\begin{align}
  E\psi (\bm R) = - \frac {\hbar^2} {2m_\alpha } \frac {\partial^2} {\partial x_\alpha^2} \psi (\bm R) + V(\bm R) \psi (\bm R)\;.
\end{align}
We write $\psi $ for the wavefunction at this resolution, which tells you how much amplitude is in each unit cell.

The potential varies with the system under consideration.
\cite{luttingerHyperfineSplittingDonor1954,kohnTheoryDonorLevels1955}, on which \cite{feherElectronSpinResonance1959} leans, consider a single donor electron (e.g. P). 
(Feher goes on to use the bunching---App.~\ref{app:bunching}---to predict the hyperfine spitting of \eqref{eq:hyperfine-fundamental} and compare to his experiments.)
We will have in mind a hard-walled box in two directions, and an infinite extent in the third.

\subsubsection{Bunching: wavefunction on scales short compared to lattice }\label{app:bunching}
The effective mass theory of tells us the amplitude in each unit cell,
but we need the amplitude at the $\ce{^{29}Si}$ site.
This is given by the full wavefunction, which we call $\psi (\bm r)$: the wavefunction that describes the electron on scales comparable to an atomic covalent radius.
We already have $\psi (\bm R)$, $\bm R$ the unit cell holding the spin impurity, from effective mass calculations.
We would expect $\psi (\bm r)$ at the impurity to scale with $\psi (\bm R)$, so define the \textbf{bunching}
\begin{align}
  \eta  = \frac{ | \psi (\bm r) |^2}{|\psi (\bm R)|^2}\;.
\end{align}
In terms of the bunching, the energy of the electron is
\begin{align}
  E &= \frac{ 2\mu _0}{3} g\mu _B\gamma _n \eta  (z_e  \sigma )|\psi (\bm R)|^2\label{eq:hyperfine-bunching}
\end{align}

The precise value of the bunching factor is in some doubt. \cite{shulmanNuclearMagneticResonance1956} measure it at
\begin{align}
  \eta  = 186 \pm 18\quad \text{[Shulman and Wyluda]};
\end{align}
this is what Feher uses.
\cite{wilsonElectronSpinResonance1964} claims that Shulman and Wyluda made two roughly countervailing errors---a factor of 2 mistake and an incorrect mobility---and quotes unpublished work of I. Solomon as giving
\begin{align}
  \eta  = 178 \quad \text{[Solomon]}.
\end{align}
(see Ref's [31, 32] of \cite{wilsonElectronSpinResonance1964}.)
Knight shift measurements of \cite{sundforsNuclearMagneticResonance1964} in phosphorus- and boron-doped Si give
\begin{subequations}
\begin{align}
  \eta  &= 100 \pm 10 \quad\text{[Sundfors; P]}\\
  \eta  &= 80 \pm 10 \quad\text{[Sundfors; B]}
\end{align}
\end{subequations}
(table V).
DFT calculations of \cite{assaliHyperfineInteractionsSilicon2011} and \cite{philippopoulosFirstprinciplesHyperfineTensors2020} give
\begin{subequations}
  \begin{align}
  \eta  &= 159.4 \pm 4.5\quad \text{[Assali et al.]}\\
     \eta  &= 88 \quad\quad\quad\qquad\text{[Philippopoulos et al.]} 
  \end{align}
\end{subequations}
We take
\begin{align}
  \eta  = 100\;.
\end{align}

\subsubsection{``Effective mass theory'' for our channel}\label{app:our-channel}

We have in mind a long wire with width $w \approx 100\;\mathrm{nm}$ and thickness $b = 10\ \text{nm}$.

To reduce the effective model to 1+1d,
factorize the (effective mass theory) wavefunction
\begin{align}
  \psi (\bm r) = \psi _x(x) \psi _y(y) \psi _z(z)
\end{align}
with $x$ along the wire and $z$ transverse to the Si layer. The $x$ and $y$ factors will be of order
\begin{subequations}
\begin{align}
  \psi _x &\sim 1/\sqrt{w}\\
  \psi _y &\sim 1/\sqrt{b}\;,
\end{align}
\end{subequations}
so the energy \eqref{eq:hyperfine-bunching} becomes

\begin{align}
  \begin{split}
    E &= \frac{ 2\mu _0}{3} g\mu _B\gamma _n \eta  (z_e  \sigma )|\psi _x|^2 |\psi _y|^2|\psi _z(0)|^2\\
    &= \frac{ 2\mu _0}{3} \frac{g\mu _B\gamma _n \eta }{wb} (z_e  \sigma )|\psi _z(0)|^2\\
  \end{split}
\end{align}
from which we can read off the parameters $au$
\begin{align}
  au =  \frac{ 2\mu _0}{3} \frac{g\mu _B\gamma _n \eta }{wb}\;.
\end{align}

\subsection{Nuclear spin contribution to the Larmor frequency landscape}\label{app:nucl-dwell-time}
The \si{29}\ nuclear spins give an Overhauser magnetic field resulting in a contribution to the electron Larmor frequency
\begin{align}
   \Omega_\text{nucl} = \sum_j \sigma_j (au/\hbar) |\psi(x_j)|^2
\end{align}
where the sum is over nuclear spin impurities.
The variance is
\begin{align}
    \sigma_{\Omega_\text{nucl}}^2 = \var \Omega_\text{nucl} \sim a^2u^2 (wb \nu) / l_e \hbar^2
\end{align}
using that $|\psi(x_j)|^2 \sim 1/l_e$ on a length scale of $l_e$,
from which we can estimate an area-corrected variance
\begin{align}
    \sigma_{0;\nucl} \sim a^2u^2w^2b\nu = \left(\frac {2\mu_0}{3} g\mu_B \gamma_n \eta/\hbar\right)^2 \frac 1 b \nu\;.
\end{align}
Returning to the results of App.~\ref{app:variable-g}, this gives 
\begin{align}
    L_2^{-1} &\sim \frac{\sigma_{0;\nucl}^2 k_BT}{el_ew\mu^2E^3} \notag\\
    &= \left(\frac {2\mu_0}{3} \frac{g\mu_B \gamma_n \eta}{\hbar \mu E}\right)^2 \frac \nu {wb} \frac{k_BT}{eEl_e}
\end{align}